\documentclass[apj]{emulateapj}
\usepackage{txfonts}
\usepackage{subfigure}
\usepackage{multirow}

\usepackage{times}
\usepackage{natbib}
\usepackage{rotating}
\usepackage{url}
\usepackage{amssymb}
\bibpunct{(}{)}{;}{a}{}{,}

\begin{document}

\title{The broad band spectral variability of MCG-6-30-15 observed by {\it NuSTAR} and {\it XMM-Newton}.}

\author{A. Marinucci\altaffilmark{1,2},  G. Matt\altaffilmark{1}, G. Miniutti\altaffilmark{2}, M. Guainazzi\altaffilmark{3}, M. L. Parker\altaffilmark{4}, L. Brenneman\altaffilmark{5}, A. C. Fabian\altaffilmark{4}, E. Kara\altaffilmark{4}, P. Arevalo\altaffilmark{6}, D. R. Ballantyne\altaffilmark{7}, S. E. Boggs\altaffilmark{8}, M. Cappi\altaffilmark{9}, F. E. Christensen\altaffilmark{10}, W. W. Craig\altaffilmark{11,12}, M. Elvis\altaffilmark{5},  C.J. Hailey\altaffilmark{12}, F. A. Harrison\altaffilmark{13}, C. S. Reynolds\altaffilmark{14}, G. Risaliti\altaffilmark{5,15}, D. K. Stern\altaffilmark{16}, D. J. Walton\altaffilmark{13}, W. Zhang\altaffilmark{17}}

\altaffiltext{1}{Dipartimento di Matematica e Fisica, Universit\`a degli Studi Roma Tre, via della Vasca Navale 84, 00146 Roma, Italy}
 \altaffiltext{2}{Centro de Astrobiolog\'ia (CSIC-INTA), Dep. de Astrofisica; ESAC, PO Box 78, Villanueva de la Ca\~{n}ada, Madrid, Spain}
\altaffiltext{3}{European Space Astronomy Center of ESA, Apartado 50727, 28080 Madrid, Spain}
\altaffiltext{4}{Institute of Astronomy, University of Cambridge, Madingley Road, Cambridge, CB3 OHA, UK}
\altaffiltext{5}{Harvard-Smithsonian Center for Astrophysics, 60 Garden Street, Cambridge, MA, USA}
\altaffiltext{6}{Pontificia Universidad Cat\'{o}lica de Chile, Instituto de Astrof\'{i}sica, Casilla 306, Santiago 22, Chile}
\altaffiltext{7}{Center for Relativistic Astrophysics, School of Physics, Georgia Institute of Technology, Atlanta, GA 30332, USA}
\altaffiltext{8}{Space Sciences Laboratory, University of California, Berkeley, 7 Gauss Way, Berkeley, CA 94720-7450, USA}
\altaffiltext{9}{INAF, IASF Bologna, Via P Gobetti 101, 40129 Bologna, Italy}
\altaffiltext{10}{Danish Technical University, DK-2800 Lyngby, Denmark}
\altaffiltext{11}{Lawrence Livermore National Laboratory, Livermore, CA, USA}
\altaffiltext{12}{Columbia University, New York, NY 10027, USA}
\altaffiltext{13}{Cahill Center for Astronomy and Astrophysics, California Institute of Technology, Pasadena, CA, 91125 USA}
\altaffiltext{14}{Department of Astronomy, University of Maryland, College Park, MD 20742-2421, USA}
\altaffiltext{15}{INAF Ð Osservatorio Astrofisico di Arcetri, L.go E. Fermi 5, I-50125 Firenze, Italy}
\altaffiltext{16}{Jet Propulsion Laboratory, California Institute of Technology, 4800 Oak Grove Drive, Pasadena, CA 91109, USA}
\altaffiltext{17}{NASA Goddard Space Flight Center, Greenbelt, MD 20771, USA}
\date{Received / Accepted}

\begin{abstract} 
MCG-6-30-15, at a distance of 37~Mpc ($z=0.008$), is the archetypical Seyfert 1 galaxy showing very broad  Fe K$\alpha$ emission.
We present results from a joint {\it NuSTAR} and {\it XMM-Newton} observational campaign that, for the first time, allows a sensitive, time-resolved spectral analysis from 0.35 keV up to 80 keV. The strong variability of the source is best explained in terms of intrinsic X-ray flux variations and in the context of the light bending model:  the primary, variable emission is reprocessed by the accretion disk, which produces secondary, less variable, reflected emission. The broad Fe K$\alpha$ profile is, as usual for this source, well  explained by relativistic effects occurring in the innermost regions of the accretion disk around a rapidly rotating black hole. We also discuss the alternative model in which the broadening of the Fe K$\alpha$ is due to the complex nature of the circumnuclear absorbing structure. Even if this model cannot be ruled out, it is disfavored on statistical grounds. We also
detected an occultation event likely caused by BLR clouds crossing the line of sight.

\end{abstract}

\keywords{Galaxies: active - Galaxies: Seyfert - Galaxies: accretion - Individual: MCG--6-30-15}

\section{Introduction}
 \begin{figure*}
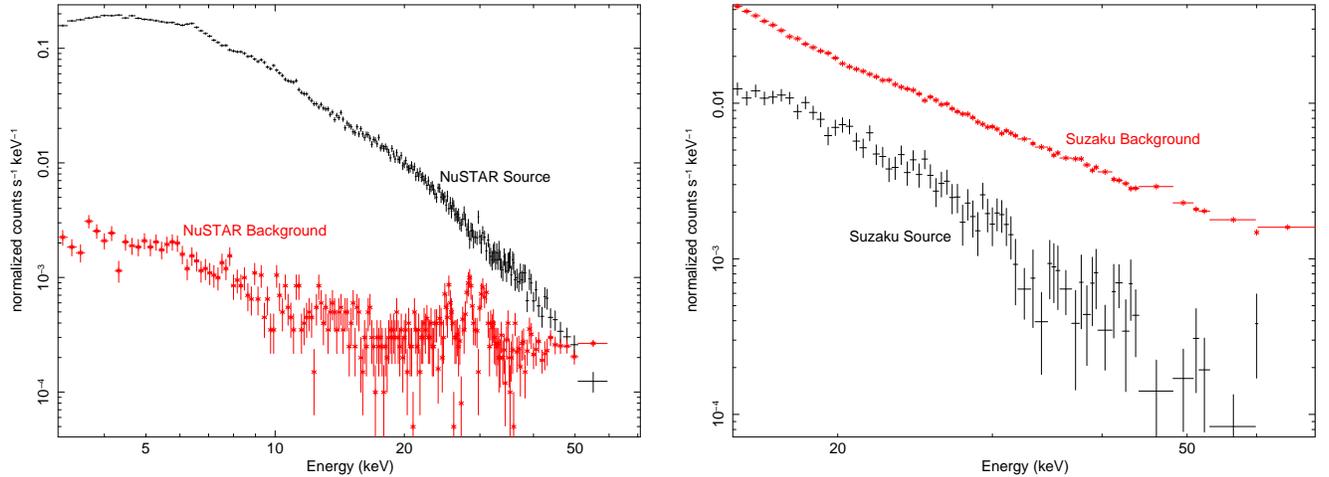

\centering
\includegraphics[width=0.78\columnwidth, angle=-90]{nustar_back.eps}
\includegraphics[width=0.78\columnwidth, angle=-90]{pin_back.eps}
\caption{\label{backgrounds} {\it Left:} source (in black) + background (in red) spectra from the {\it NuSTAR} FPMA in the 3-80 keV band. {\it  Right:} archival {\it Suzaku} HXD-PIN source (in black) + background (in red) spectra in the 15-70 keV band. The source is at the same 15-70 keV flux level in both observations, within a few per cent.}
\end{figure*}

The bright Seyfert 1 galaxy MCG-6-30-15 ($z$=0.00775) is the first source in which a broad iron K$\alpha$ line was detected with {\it ASCA }\citep{tanaka95}, showing a red tail whose low energy extension 
is an indicator of the inner radius of the accretion disk and thus of the black hole spin \citep{iwa96,ify99}. The iron K$\alpha$ line is very prominent in this source, since the iron abundance appears to be significantly
higher than solar \citep{fab02}. Due to its spectroscopic features, MCG-6-30-15 is one of the most observed AGN in the X-rays. It was observed several times with {\it ASCA} \citep{shih02, mif03}, {\it BeppoSAX} \citep{gmmo99}, {\it RXTE} \citep{lee99, ve01}, {\it XMM-Newton} \citep{wilms01,fab02, fv03, vf04, br06} and {\it Suzaku} \citep{miniutti07, noda11}; multi-observatory data has also been analyzed by \citet{mtr08} and \citet{chfa11}. The soft X-ray spectrum of this source has a complex structure due to warm absorption \citep{otani96}. It has been studied also at high resolution with the {\it Chandra} HETGs \citep{lee01, ylf05} and {\it XMM-Newton} RGS \citep{bsk01}. \citet{tfv03, tflv04} confirmed the presence of dusty warm absorbers, in agreement with optical observations \citep{rey97b}. 

The extreme variability of MCG-6-30-15 in the X-rays has often been explained with a scenario where two components play the major role: a highly variable power law continuum (with an almost constant photon index) and a much less variable reflection spectrum from the innermost region of the accretion disk (within a few gravitational radii) \citep{shih02, fv03, tum03, miniutti07, parker13}.  
The light-bending model \citep{fv03, min03, mf04}, a generalization of earlier work \citep{mama96, rb97}, attributes the change of the power law flux to the variation of the location of the X-ray emitting source close to the central black hole. In this scenario much of the radiation is bent down onto the disk and the observed variation in the reflection intensity is small because a large fraction of photons does not escape to infinity but is instead captured by the black hole. The detection of a strong reflection hump, peaking at $\sim$30 keV, in previous high energy observations of MCG-6-30-15 by {\it BeppoSAX} \citep{gmmo99}, {\it RXTE} \citep{lfr00} and {\it Suzaku} \citep{miniutti07} is consistent with this two-component model. 

An alternative absorption-dominated model has also been used to explain the extreme behavior of MCG-6-30-15 \citep{mtr08, mtr09}. In this model the red wing of the line is not due to strong relativistic effects but to the complex structure of absorbers along the line of sight \citep{miller07, turner07}. These complex absorbing structures (with column densities in the $10^{22}$--$10^{24}$ cm$^{-2}$ range) can produce an apparent broadening of the Fe K$\alpha$ emission line by partially covering the nuclear X-ray source. The covering factor of some of the obscuring media may need to be linked to variations in the nuclear flux, as already shown in the past for the case of MCG-6-30-15 \citep{mtr08}. This interpretation ascribes the constancy of the amplitude of the iron line to the greater distance of the emitting material from the variable X-ray source, while the hard flux excess above $\sim$20 keV is interpreted as originating from Compton-thick clouds at or within the Broad Line Region, partially covering the X-ray nuclear source \citep{tatum13}.

 We present results from a simultaneous {\it NuSTAR} and {\it XMM-Newton} observational campaign performed in January 2013. Taking advantage of the unique {\it NuSTAR} high-energy sensitivity, we simultaneously cover the 0.35--80 keV energy band with unprecedented signal to noise ratio. The primary focus of this paper is the spectral variability of this source, and understanding how the spectral components vary. We  discuss the results in the context of the two scenarios described above. The paper is structured as follows: in Sect. 2 we discuss the joint {\it NuSTAR} and \textit{XMM-Newton} data reduction, in Sect. 3, 4 and 5 the spectral analysis and best fit parameters are presented and discussed within a reflection and absorption scenario, respectively. Sect. 6 is devoted to the spectral variability by occultation from Broad Line Region clouds and Sect. 7 to the flux-flux plots.

\section{Observations and data reduction}
\subsection{\it NuSTAR}
{\it NuSTAR}  \citep{nustar}  observed MCG-6-30-15 simultaneously with {\it XMM-Newton} with its two coaligned telescopes with corresponding Focal Plane Modules A (FPMA) and B (FPMB) starting on 2013 January 29 for a total of $\sim 360$ ks of elapsed time. The Level 1 data products were processed with the {\it NuSTAR} Data Analysis Software (NuSTARDAS) package (v. 1.1.1). Event files (level 2 data products) were produced, calibrated, and cleaned using standard filtering criteria with the \textsc{nupipeline} task and the latest calibration files available in the {\it NuSTAR} calibration database (CALDB). Both extraction radii for the source and background spectra were 1.5 arcmin. Spectra were binned in order to over-sample the instrumental resolution by at least a factor of 2.5 and to have a Signal-to-Noise Ratio (SNR) greater than 3$\sigma$ in each spectral channel. Exposure times and total counts for each spectrum can be found in Table \ref{extraction}.

Figure \ref{backgrounds} shows a comparison between the non-imaging {\it Suzaku} HXD-PIN spectrum analyzed in \citet{miniutti07} and \citet{noda11} and the {\it NuSTAR} FPMA spectrum. The low background of  {\it NuSTAR} above 10 keV is unprecedented. The two spectra have the same net exposure times ($\sim 120$ ks) and average 15-70 keV flux ($6.5\times 10^{-11}$ ergs cm$^{-2}$ s$^{-1}$), within a few percent. Yet in  {\it NuSTAR}  the ratio of the source to background at 20 keV is $\sim25$, while it is $\sim0.25$ in \emph{Suzaku}: the factor 100 gain is due to the {\it NuSTAR} focusing optics.

\subsection{\it XMM-Newton}
\begin{figure*}[t!]
\centering
\includegraphics[width=\textwidth]{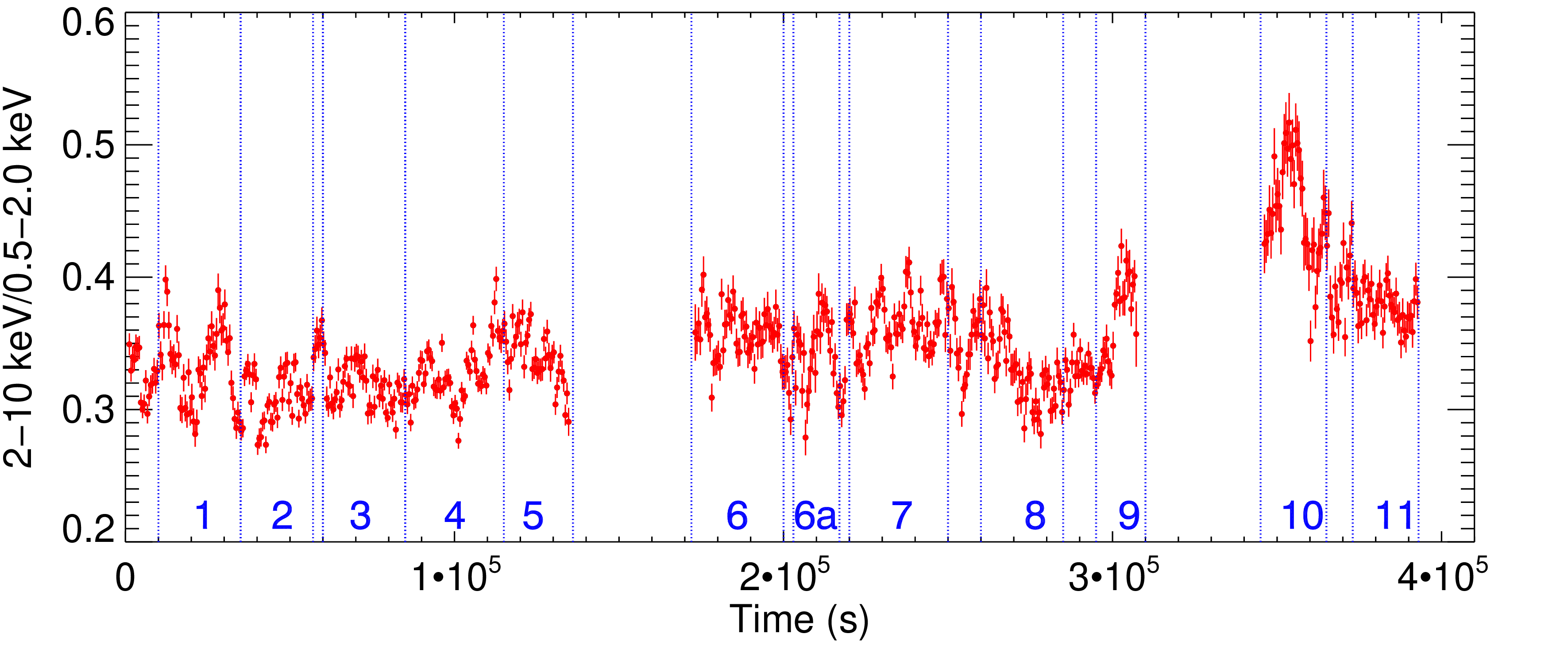}
\caption{\label{intervals} Ratio between the 2-10 keV and 0.5-2.0 keV light curves (in 500 s bins) and time intervals chosen for our analysis. Data are from {\it XMM-Newton} EPIC-Pn camera only and time is from the start of the observation. }
\end{figure*}

MCG-6-30-15 was observed by \textit{XMM-Newton} \citep{xmm} for $\sim$315 ks starting on 2013 January 29 during three consecutive revolutions (OBSID 0693781201, 0693781301 and 0693781401) with the EPIC CCD cameras, the Pn \citep{struder01} and the two MOS \citep{turner01}, operated in small window and medium filter mode. The three EPIC-Pn event files were merged with the ftool \textsc{merge} into one single event file. Data from the MOS detectors are not included in our analysis since they strongly suffered from photon pileup. The extraction radii and the optimal time cuts for flaring particle background were computed with SAS 12 \citep{gabr04} via an iterative process which leads to a maximization of the SNR, similar to that described in \citet{pico04}. The resulting optimal extraction radius is 40 arcsec and the background spectra were extracted from source-free circular regions with a radius of about 50 arcsec.
\begin{figure*}[t!]
\centering
\includegraphics[width=\textwidth]{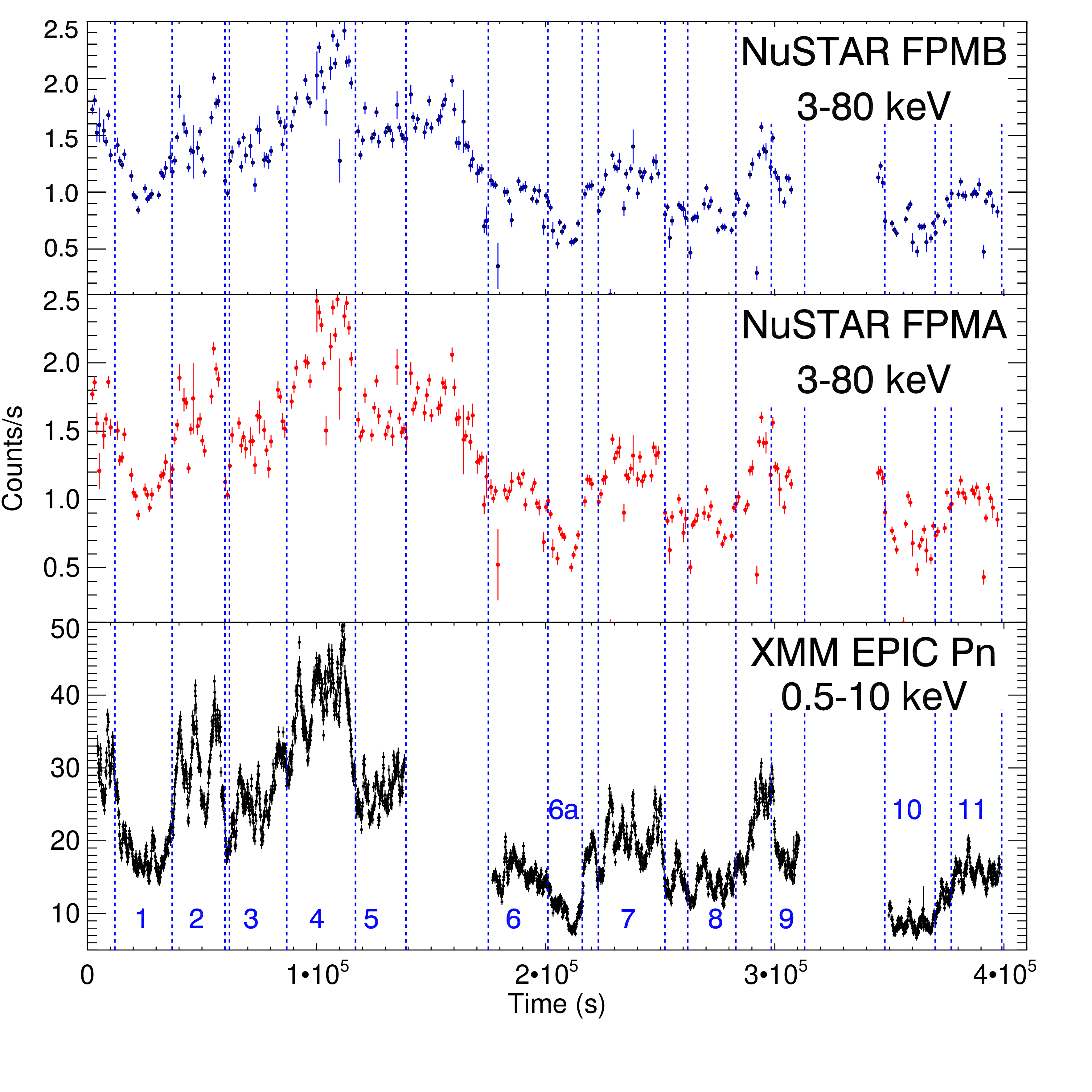}\\
\caption{\label{lc}  From the top to the bottom, {\it NuSTAR} FPMB and FPMA (in the 3-80 keV energy interval) and {\it XMM-Newton} EPIC-Pn (in the 0.5-10 keV energy interval) light curves. Count rate for the instruments is plotted versus the time from the start of the observation; vertical lines indicate the different time intervals used in our analysis.}
\end{figure*}
In Figure \ref{lc} (bottom panel) the 0.5-10 keV light curve of the source can be seen and we get average count rates of $28.316\pm0.018$, $16.756\pm0.014$ and $12.181\pm 0.020$ counts/s for the three orbits, respectively. The source is highly variable both in flux and in spectral shape: applying cuts only in flux could mix different spectral states. Hence, spectra were extracted from 11 intervals with the aim of choosing states with constant hardness ratio (Figure \ref{intervals}). Details on net exposure times and total counts can be found in Table \ref{extraction}. 
Spectra were binned in order to over-sample the instrumental resolution by at least a factor of 3 and to have no less than 30 counts in each background-subtracted spectral channel. This allows the application of $\chi^2$ statistics. We do not include the 2.0-2.5 keV energy band in our analysis due to instrumental effects that are discussed in Appendix A.1.

Due to the well-known extreme variability of the source choosing strictly simultaneous data is essential. Using the ftool \textsc{mgtime} we merged the good time intervals tables of the two telescopes and only simultaneous observing windows are used in the following analysis.

The RGS spectra were reduced following the guidelines in \citet{guabia07}. We used the data reduction pipeline \textsc{rgsproc}, coupled with the latest calibration files available. We chose a fixed celestial reference point for the attitude solution, coincident with the NED optical nucleus of MCG-6-30-15. Source spectra were extracted in regions of the dispersion versus cross-dispersion and Pulse Invariant versus cross-dispersion planes, corresponding to 95\% of the Point Spread Function (PSF) in the cross-dispersion direction. Background spectra were generated using a sub-set of blank field observations whose background counts matched the level measured during each individual RGS observation. Using the ftool \textsc{rgscombine} we obtained 315 ks of net exposure time. \\

\begin{table}
\caption{\label{extraction} Net exposure times and total counts for the data sets used in this work. Epic-Pn count rates are calculated in the 0.5-10 keV energy band while for the FPMA and FPMB detectors we used the 3-80 keV band.}
\begin{center}
\scriptsize
\begin{tabular}{c|cc|ccc}
{\bfseries Interval } & {\bf Exp. time (ks)}& & &{\bf Counts}  \\
\hline
& EPIC Pn & FPMA, FPMB& EPIC Pn & FPMA & FPMB\\
\hline
1 & 10.4&13.6&166492 & 15383&14563\\
2& 8.5& 8.9 & 201311& 13255&12647\\
3& 7.5&9.6 & 180440& 13772&12974\\
4&10.8& 13.6&249127 & 27595&25925\\
5&8.6 & 10.4& 169247& 15821&15440\\
6& 10.3& 14.0& 132764& 14018&13324\\
7& 11.1& 14.7& 170145&16683 &15675\\
8& 7.9& 10.7& 97936&8825 &8535\\
9& 4.7& 6.0& 76363& 7115&6683\\
10&6.5 & 9.0& 48437& 6636&6291\\
11& 7.7& 12.5&99878 & 12288&11831\\
\hline
\hline
\end{tabular}
\end{center}
\end{table}

\section{ Data analysis}
\begin{figure}[t!]
\centering
\includegraphics[width=0.75\columnwidth, angle=-90]{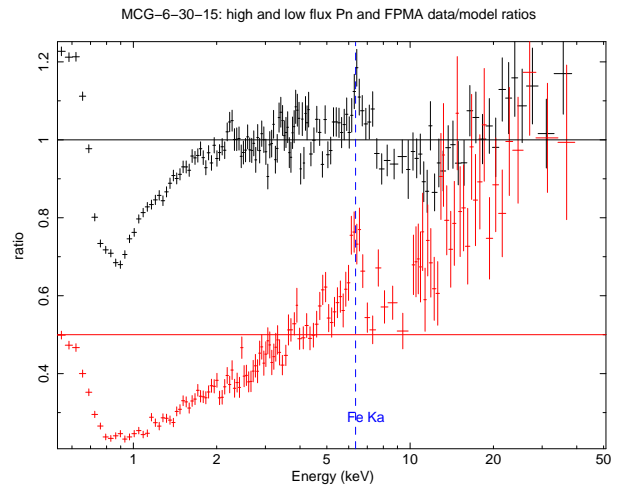}\\
\caption{\label{ratios} EPIC-Pn (0.5-10 keV) and FPMA (above 10 keV) ratios to a $\Gamma=2$ power law, rescaled for a normalization factor. Top spectra (in black) have been extracted from interval 4 (high flux state) and bottom spectra have been extracted from interval 10 (low flux state). A strong reflection hump and broad Fe K$\alpha$ line are present in the low flux state. In the high flux state (black) the effect of the warm absorbing structure can be clearly seen below 3 keV, while the effect of occultation by BLR clouds is present in the low flux state (red). Some binning is applied for the sake of clarity in the four spectra.}
\end{figure}

Figure \ref{ratios} shows the ratios of {\it NuSTAR} and {\it XMM-Newton} data sets to a  $\Gamma=2$ power law in a high (\#4) and low (\#10) flux state, with the aim of identifying the different features we will consider in our spectral analysis. The most important features above 3 keV are the large Compton hump peaking around 20-30 keV,  indicating the important role of reflection above 10 keV \citep{gf91, faro10} and the broad Fe K$\alpha$ line that has been extensively studied in the past.  At softer energies (below 3 keV) we see features from a complex ionized absorber \citep{lee01, sako03} and a soft excess below $\sim$0.7 keV that is frequently seen in AGN \citep{gido04,crummy06,mipogre09,wana13}. \\
However, as the RGS data are not well suited to constrain a highly complex continuum model, we first consider a simplified, phenomenological continuum model comprising only a power law and black body components.
Our analysis procedure used the following strategy: we first identify and characterize the warm absorbing structure taking advantage of the high resolution RGS spectra with the continuum inferred in \citet{lee01}.  Once we reach a satisfactory fit, we apply this model component to the broad band spectral analysis. As a cross check, we load the broadband best fit model for the continuum back into the RGS spectra to calculate errors and final best fit parameters. \\
\subsection{\label{rgs}RGS spectral analysis}
We fit the spectra of MCG-6-30-15 with a model consisting of the following components: reflection from cold matter distant from the central X-ray source; relativistically blurred reflection from an ionized accretion disk; and a redshifted power law for the primary nuclear emission. We used \textsc{xillver} for both the cold and ionized reflection \citep{garcia13} and \textsc{relconv} for relativistic smearing \citep{dauser13}. 
The three components described above were then convolved with two ionized absorbers, one dusty absorber and Galactic absorption \citep[\textsc{tbabs}, 3.92$\times10^{20}$ cm$^{-2}$;][]{dl90}. Tables for the ionized absorbers were generated using \textsc{xstar v.2.2.0}. The source luminosity between 1 and 1000 Ryd was assumed to be $10^{44}$ erg s$^{-1}$ with a powerlaw spectrum with $\Gamma=2.0$, the turbulent velocity was set to 200 km s$^{-1}$, the density to 10$^{12}$ cm$^{-3}$, the temperature to $10^4$ K, and the covering factor to 1. We refer the reader to \citet{lee01} for further details about absorption by dust in MCG-6-30-15. A cross-calibration constant has been left free to vary when fitting FPMA, FPMB and Pn spectra simultaneously.

\begin{figure*}
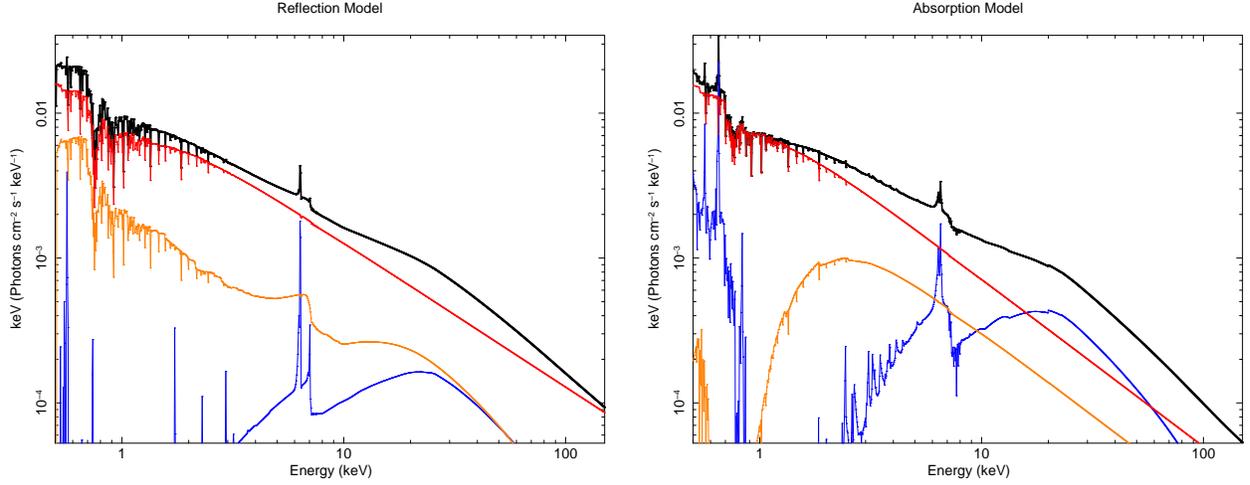

\centering
\includegraphics[width=0.75\columnwidth, angle=-90]{refl6_mo.eps}
\includegraphics[width=0.75\columnwidth, angle=-90]{abs6_mo.eps}
\caption{\label{models} Theoretical models used in this work. Both left and right hand plots are best fitting models for spectra extracted in the same time interval. {\it Left:} the reflection model is plotted in the 0.5-150 keV energy band with the following main components: the primary continuum (in red), blurred ionized disk reflection (in orange) and cold, distant reflection (in blue). {\it Right:} the absorption model is plotted in the 0.5-150 keV energy band with the following main components: the primary unabsorbed (in red) and absorbed (in orange) power laws and reflection from distant, ionized material (in blue). }
\end{figure*}

 In \textsc{xspec} the model reads as follows: \\\\
\textsc{tbabs}$\times$\textsc{warmabs$_1$}$\times$\textsc{warmabs}$_2\times$ \textsc{dustyabs}$\times$\\
(\textsc{xillver$_1$}+\textsc{relconv}$\times$\textsc{xillver$_2$}+ \textsc{zpow})\\\\
and be seen in Figure \ref{models} (left panel).
The adopted cosmological parameters are $H_0=70$ km s$^{-1}$ Mpc$^{-1}$, $\Omega_\Lambda=0.73$ and $\Omega_m=0.27$, i.e. the default ones in \textsc{xspec 12.8.1} \citep{xspec}. Unless otherwise stated, errors correspond to the 90\% confidence level for one parameter of interest ($\Delta\chi^2=2.7$). The RGS spectra were re-binned only for clarity in the plots and were analysed using Cash statistics \citep{cash76}.

 \begin{figure*}
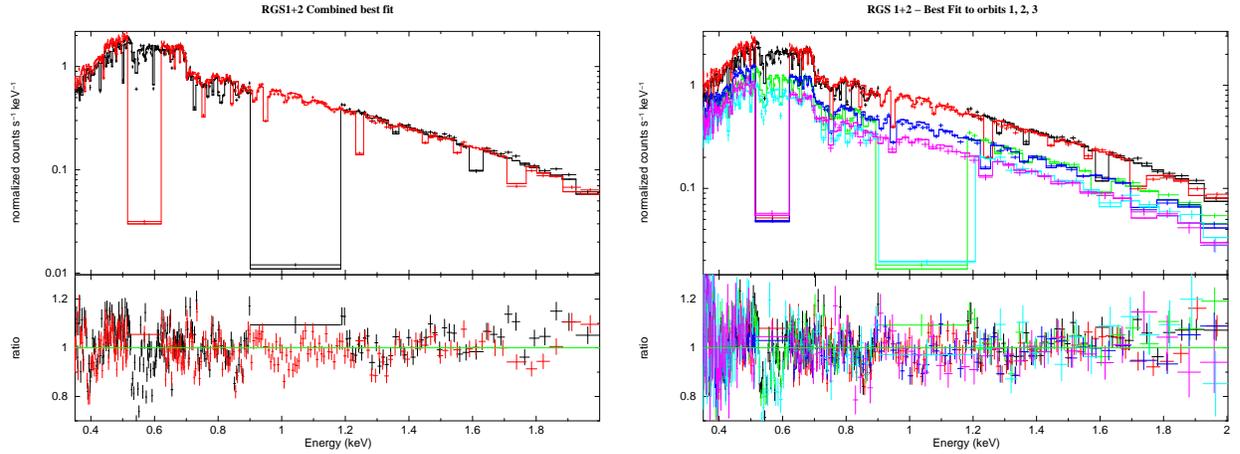

\centering
\includegraphics[width=0.7\columnwidth, angle=-90]{rgs_combined.eps}
\includegraphics[width=0.7\columnwidth, angle=-90]{plot_3rgs.eps}
\caption{\label{rgs_combined} \textit{Left:} RGS1+2 best fit. Spectra are obtained combining data from the three separate orbits. \textit{Right:} RGS1+2 best fits from the three separate observations. Spectra indicate, from the top to the bottom, high flux (first orbit), medium flux (second orbit) and low flux (third orbit) states, respectively. Gaps around 0.5 keV and 1.0 keV indicate instrumental effects due to the loss of CCDs in the RGS detectors.}
\end{figure*}
\begin{table}
\caption{\label{rgs_bestfit} RGS1+2 best fit parameters. Column densities are in $10^{21}$ cm$^{-2}$ units, ionization parameters $\xi$ are in erg cm s$^{-1}$ units.}
\begin{center}
\scriptsize
\begin{tabular}{cccccc}
{\bfseries Parameter } & Combined &Orbit 1& Orbit 2&  Orbit 3 \\
\hline
\hline
 N$_{\rm H1}$ &$8.9\pm0.5$&   $8.2_{-0.9}^{+0.6}$ & $8.5_{-2.4}^{+0.9}$ & $5.5_{-2.0}^{+2.3}$\\
 $\log(\xi_1)$&$2.02\pm0.01$&  $2.03\pm0.01$ &$2.03\pm0.01$&$2.27_{-0.12}^{+0.18}$ \\
 N$_{\rm H2}$&$0.9\pm0.1 $&  $0.8\pm0.1 $  &  $0.9_{-0.2}^{+0.5}$  &  $3.9_{-1.0}^{+1.2}$  \\
 $\log(\xi_2)$& $1.43\pm0.05$&   $1.39\pm0.07$&$1.50_{-0.14}^{+0.07}$  & $1.77\pm0.04$ \\
 $\log (\rm N_{\rm Fe})$&$17.37\pm0.05$  &17.37*& 17.37*&17.37*\\
  & & & & & \\
$\Gamma$& $2.03\pm0.01$ &2.03 *  &2.03*&2.03*        \\
 N$_{\rm pow} (\times10^{-2})$&$1.50\pm0.02$    &  $2.11\pm0.03$&$1.26\pm0.04$& $0.86\pm0.03$  \\
 $\log(\xi_{\rm refl})$&$1.34\pm0.03$  &    $1.50_{-0.07}^{+0.11}$   & $1.09_{-0.06}^{+0.10}$ & $0.70_{-0.32}^{+0.09}$   \\
  N$_{\rm Refl}(\times10^{-5})$&$1.8\pm0.2$ & $1.5\pm0.3$&$2.2\pm0.5$&$7.0\pm{2.2}$\\
  \hline
  \hline
\end{tabular}
\end{center}
\end{table}

RGS1+2 combined spectra were first fitted with a model consisting of a power law ($\Gamma=1.84$) and a black body   \citep[kT=0.13 keV,][]{lee01}. The fit is poor (C/dof=27750/4912) and the inclusion of a warm absorber (N$_{\rm H}=1.8\pm0.4\times10^{22}$ cm$^{-2}$, $\log(\xi / ({\rm erg}\, {\rm cm}\, {\rm s}^{-1}))=1.964\pm0.002$) strongly improves the fit (C/dof=13154/4910). The addition of a second absorber further improves it (C/dof=10687/4908) and a further, dusty absorber ($\log N_{\rm Fe}=17.37\pm0.05$) is also required ($\Delta$C=1592). 
The final best fit (C/dof=9095/4907) parameters for the warm absorbers are in Table \ref{rgs_bestfit}.
 A marginal improvement in the fit ($\Delta$C=35) is found adding a third absorber to the model (N$_{\rm H3}=2.6^{+2.0}_{-1.0}\times10^{21}$ cm$^{-2}$, $\log(\xi_3 / ({\rm erg}\, {\rm cm}\, {\rm s}^{-1}))=2.7\pm0.2$) with no variations of the other absorbers. As this component is not required we omit it from further consideration. When we leave the outflowing velocities free to vary no improvement is found.
 
  We then applied this model for the absorbing structure to the broadband spectra  (detailed discussion in Sect. \ref{fulldataset}). Once a best fit is obtained for the 11 time intervals, we removed the power law and the black body component in the RGS fit and introduced the continuum inferred from the joint FPMA-FPMB-Pn analysis. The new best fit parameters for the warm absorbers are consistent with the ones discussed above.  The best fit (Figure \ref{rgs_combined}, left panel) leads to a C/dof=8936/4905 and no additional components are needed to model the underlying continuum.
  
We then applied the best fit model to the six RGS spectra extracted from the three separate {\it XMM} orbits,  in order to search for variations in the warm absorbing material on long timescales. We get a good fit leaving the normalizations of the power laws free (C/dof=19440/14626) and if we allow the parameters to vary between the three sets of data we find a best fit (Figure \ref{rgs_combined}, right panel) of  C/dof= 19036/14619 with no significant variation of the warm absorbers (Table \ref{rgs_bestfit}; throughout the text parameters with asterisks indicate fixed values). 

We note that residuals between 0.5 and 0.6 keV in the six spectra can be ascribed to to Galactic absorption lines (OI at 0.527 keV) and to further absorption lines at the redshift of the source, which do not affect the broad band best fit values. A detailed, high resolution model of the warm absorbing structure in MCG-6-30-15 is beyond the scope of this work.\\

\begin{table*}
\caption{\label{refl_best_fit} Best fit parameters for the reflection model.\\ Column densities are in $10^{22}$ cm$^{-2}$ units, ionization parameters $\xi$ are in erg cm s$^{-1}$ units.
Columns: (a) ionization parameter of the first warm absorber; 
(b) Column density of the first warm absorber; 
(c) Ionization parameter of the second warm absorber;
(d) Column density of the second warm absorber;
(e) Iron column density of the dusty absorber;
(f) Normalization of the neutral reflection component ($\log (\xi)=0$);
(g) Ionization parameter for the reflection component from the accretion disk;
(h) Iron abundance with respect to the solar value;
(i) Emissivity index q ($\epsilon(r)\sim r^{-q}$);
(l) Normalization of the ionized reflection component;
(m) Photon index of the primary power law;
(n) Normalization of the primary continuum component.}
\begin{center}
\scriptsize	
\begin{tabular}{c|cc|cc|c|c|cccc|cc}
 &  $\log (\xi_1)$& ${\rm N_{H_1}}$ & $\log (\xi_2)$ &${\rm N_{H_2}}$& $\log ({\rm N_{Fe}})$&N$_1$ ($\times 10^{-4}$)&$\log (\xi_{\rm refl})$ & ${\rm A_{Fe}}$& q&N$_2$ ($\times10^{-5}$)&$\Gamma$& N$_3$ $(\times10^{-2})$\\
 {\bfseries Interval }& (a) & (b) & (c) & (d) & (e) & (f) & (g) & (h) & (i) & (l) & (m) & (n) \\
 \hline
 \hline
    & & & & & & & & & & &\\
 1 &$1.98^{+0.02}_{-0.02}$ & $1.60^{+0.10}_{-0.10}$&$<0.6$ & $0.07^{+0.02}_{-0.02}$& $16.83_{-0.16}^{+0.10}$& $1.2\pm0.2$&$2.76_{-0.06}^{+0.07}$ &$2.21_{-0.17}^{+0.23}$&$2.95\pm0.15$& $0.038\pm0.008$&$2.061\pm0.005$&$1.38\pm0.05$\\
   & & & & & & & & & & &\\
 2 & $2.00^{+0.03}_{-0.02}$& $1.00^{+0.21}_{-0.22}$& $1.46_{-0.15}^{+0.10}$&$0.35^{+0.13}_{-0.13}$ & $16.83^*$&$1.2^*$ & $2.86_{-0.10}^{+0.12}$&$2.21^*$&$2.95^*$&$0.037\pm0.009$&$2.061^*$ &$1.88\pm0.09$\\
    & & & & & & & & & & & \\
  3 & $2.05^{+0.05}_{-0.04}$&$0.60^{+0.19}_{-0.15}$ &$1.47_{-0.12}^{+0.06}$ &  $0.55^{+0.15}_{-0.09}$&$16.83^*$ & $1.2^*$&  $2.87_{-0.09}^{+0.12}$&$2.21^*$& $2.95^*$&$0.041\pm0.008$&$2.061^*$&$1.73\pm0.09$\\
     & & & & & & & & & & &\\
   4 &$2.03^{+0.04}_{-0.03}$ & $0.72^{+0.15}_{-0.16}$& $1.22_{-0.09}^{+0.09}$& $0.32^{+0.08}_{-0.09}$&$16.83^*$ &$1.2^*$& $2.98_{-0.17}^{+0.02^p}$&$2.21^*$& $2.95^*$& $0.024\pm0.006$&$2.061^*$& $2.70\pm0.08$\\
      & & & & & & & & & & &\\
   5 & $1.96^{+0.03}_{-0.03}$& $0.91^{+0.16}_{-0.13}$ &  $1.15_{-0.09}^{+0.15}$&$0.13^{+0.11}_{-0.08}$  &$16.83^*$ & $1.2^*$& $2.08_{-0.09}^{+0.15}$&$2.21^*$& $2.95^*$&$0.20\pm0.05$&$2.061^*$& $2.09\pm0.12$\\
      & & & & & & & & & & &\\
    6 &$1.97^{+0.03}_{-0.04}$ &  $1.06^{+0.14}_{-0.18}$& $1.65_{-0.22}^{+0.15}$& $0.24^{+0.15}_{-0.09}$ & $16.83^*$&  $1.2^*$&$0.36_{-0.12}^{+0.23}$ & $2.21^*$&$2.95^*$&$13.0\pm3.0$&$2.061^*$& $1.22\pm0.06$ \\
       & & & & & & & & & & &\\
     7 & $1.99^{+0.06}_{-0.05}$ &$1.07^{+0.13}_{-0.15}$&$1.19_{-0.08}^{+0.13}$ & $0.17^{+0.07}_{-0.14}$& $16.83^*$& $1.2^*$&$0.76_{-0.22}^{+0.27}$ & $2.21^*$& $2.95^*$&  $3.8\pm0.9$&$2.061^*$ & $1.54\pm0.07$\\
        & & & & & & & & & & &\\
  8 & $1.98^{+0.07}_{-0.08}$& $1.02^{+0.13}_{-0.14}$&  $1.62_{-0.28}^{+0.09}$& $0.27^{+0.15}_{-0.11}$& $16.83^*$& $1.2^*$& $1.80_{-0.18}^{+0.31}$& $2.21^*$&$2.95^*$&$0.26\pm0.07$&$2.061^*$& $1.08\pm0.05$\\
     & & & & & & & & & & &\\
  9 & $1.97^{+0.06}_{-0.06}$&$1.02^{+0.16}_{-0.15}$ & $1.37_{-0.18}^{+0.15}$&  $0.22^{+0.13}_{-0.10}$& $16.83^*$&$1.2^*$&$1.53_{-0.21}^{+0.26}$ &$2.21^*$&$2.95^*$& $0.72\pm0.16$&$2.061^*$&$1.56\pm0.08$\\
     & & & & & & & & & & &\\
  10 & $1.95^{+0.02}_{-0.03}$& $2.25^{+0.15}_{-0.15}$  & $1.47_{-0.33}^{+0.42}$& $< 0.09$& $16.83^*$& $1.2^*$ &$0.08_{-0.07}^{+0.10}$ &$2.21^*$& $2.95^*$&$34.0\pm8.0$&$2.061^*$&$0.77\pm0.04$\\
     & & & & & & & & & & &\\
  11 & $2.01^{+0.03}_{-0.05}$& $0.55^{+0.21}_{-0.15}$& $1.27_{-0.13}^{+0.22}$& $0.35^{+0.11}_{-0.17}$&$16.83^*$ & $1.2^*$& $0.81_{-0.27}^{+0.31}$& $2.21^*$&$2.95^*$&$50.0\pm12.0$&$2.061^*$ &$1.24\pm0.06$\\
    & & & & & & & & & & &\\
\hline
\hline
\end{tabular}\\
\end{center}
\end{table*}
 \begin{figure*}
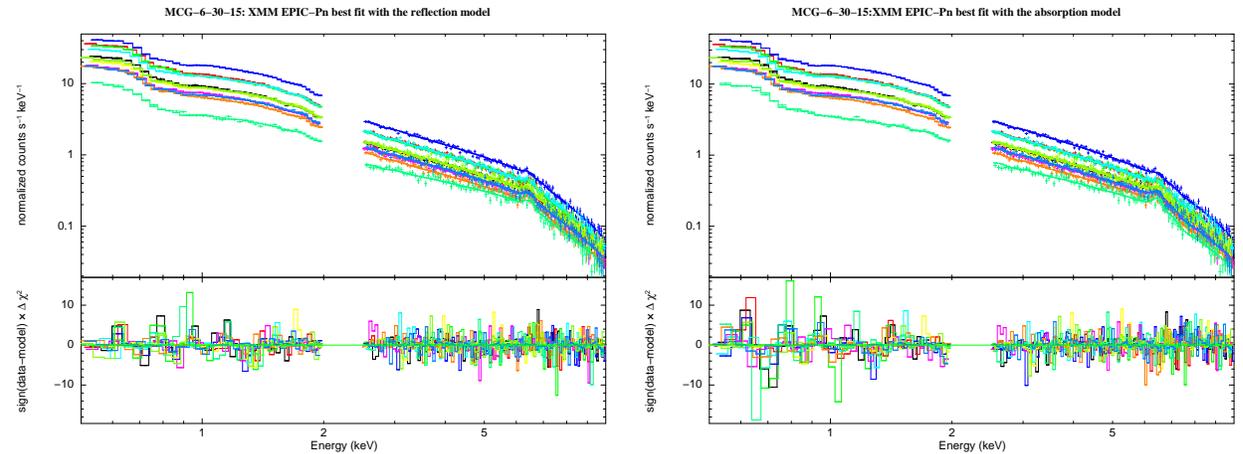

\centering
\includegraphics[width=0.7\columnwidth, angle=-90]{XMM_fit.eps}
\includegraphics[width=0.7\columnwidth, angle=-90]{abs_XMM_FIG.eps}
\caption{\label{xmmfit} Best fit and residuals of the 11 {\it XMM} EPIC-Pn spectra with the reflection (left panel) and absorption model (right panel). Gaps around 2 keV are due to EPIC-Pn calibration effects and are discussed in Appendix A.1.}
\end{figure*}

\begin{figure*}
\centering
\includegraphics[width=\textwidth]{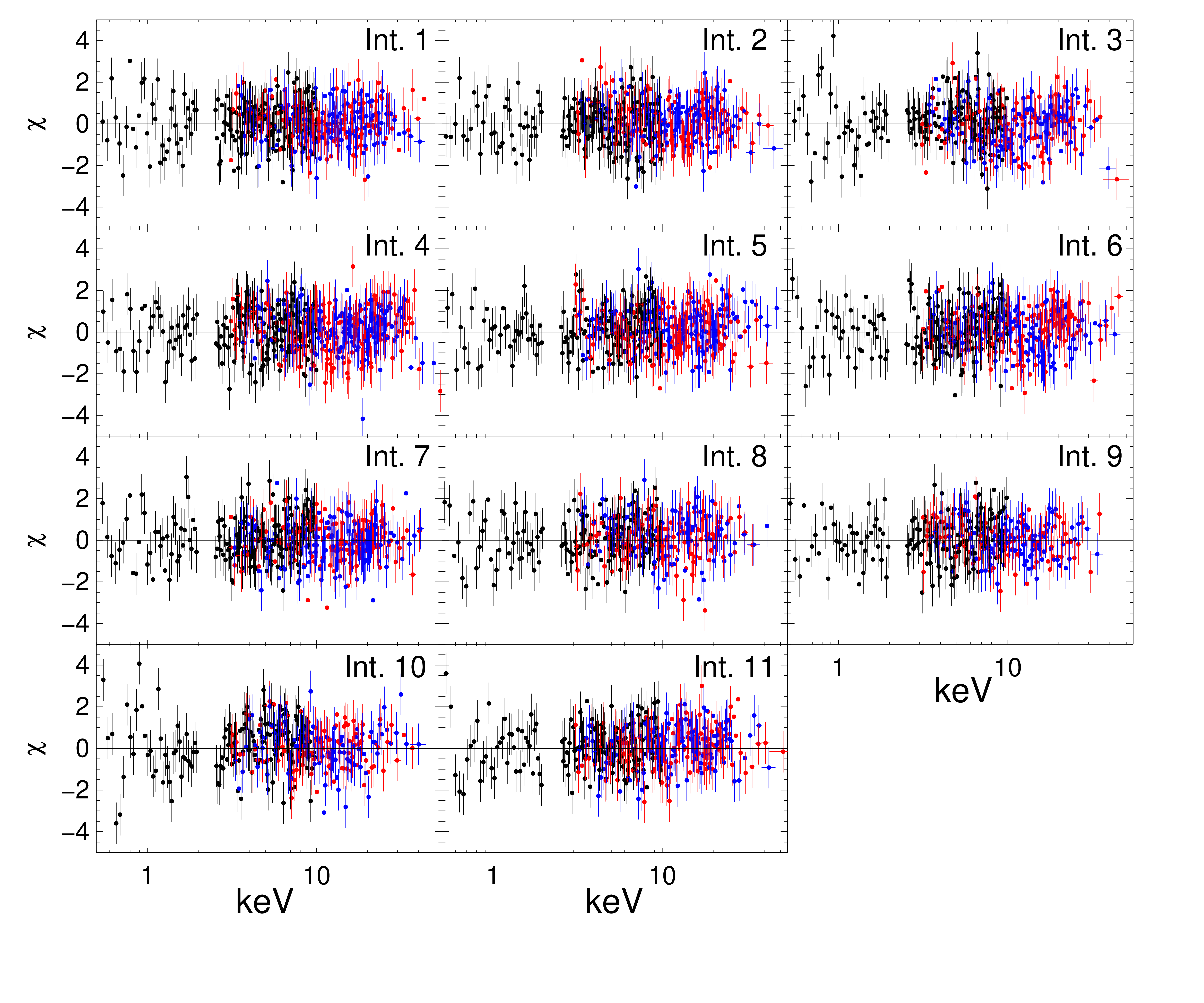}
\caption{\label{reffit} Residuals of the 33 {\it XMM} EPIC-Pn (in black) and {\it NuSTAR} FPMA (in red), FPMB (in blue) spectra with the reflection model. Gaps around 2 keV are due to EPIC-Pn calibration effects and are discussed in Appendix A.1.}
\end{figure*}

\section{The reflection scenario}\label{analysis}
We then applied the model presented in Sect. \ref{rgs} to the 11 EPIC-Pn spectra, searching for variations on shorter timescales. All parameters are tied together in the fit, with the exception of the normalization of the primary power law, the disk reflection component, and its ionization state which we allow to vary between the 11 intervals. We get a $\chi^2$/dof=2078/1595=1.30 with some residuals present, indicating a more complex interplay between the parameters. We note a significant variation of the ionization parameter of the disk reflection component among the observations. The best fit parameters for the combined {\it XMM} fit are N$_{\rm H1}=8.3\pm0.5\times10^{21}$ cm$^{-2}$, $\log(\xi_1 / ({\rm erg}\, {\rm cm}\, {\rm s}^{-1}))=1.99\pm0.01$ and N$_{\rm H2}=1.5\pm0.3\times10^{21}$ cm$^{-2}$, $\log(\xi_2 / ({\rm erg}\, {\rm cm}\, {\rm s}^{-1}))=1.33\pm0.10$ for the first and second warm absorber, respectively. 
We then left the values for the warm absorbing structure free to vary between the 11 spectra and we get a significant improvement of the fit ($\Delta\chi^2$=194) with a combined normalization of the cold reflection of N$_1=1.2\pm0.2\times 10^{-4}$. If we leave the last parameter free to vary between the 11 intervals a  marginal improvement of the fit is found ($\chi^2$/dof=1859/1545=1.20) and no strong residuals are present (Figure \ref{xmmfit}, left panel). On the other hand, when we leave the ionization state of this second reflector free no improvement in the fit is found and only an upper limit can be measured $\log(\xi / ({\rm erg}\, {\rm cm}\, {\rm s}^{-1}))<0.2$, consistent with the value found by \citet{bvf03}. We then inferred the flux of the narrow component of the iron K$\alpha$ line by measuring the flux of the neutral \textsc{xillver} component between 6.35 and 6.45 keV. We get a value of $2.0\pm0.5$ ph cm$^{-2}$ s$^{-1}$, in agreement with previous {\it Chandra} HETGs measurements \citep{lih02, yaq04}.

Variations of the warm absorbing material are found with respect to the combined best fit value in interval 1 and 10, which we discuss in Sect. \ref{reflres} and \ref{eclipses}. It is worth noting that no iron XXV K$\alpha$ or iron XXVI K$\alpha$ absorption lines are detected in the 2013 observations; however, the addition of an emission line at $6.60\pm0.05$ keV with a flux of $1.1\pm0.3\times 10^{-5}$ ph cm$^{-2}$ s$^{-1}$ improves the fit ($\chi^2$/dof=1796/1543=1.17).

\subsection{Broadband spectral analysis}\label{fulldataset}
We then introduced the 11 pairs of spectra from {\it NuSTAR}-FPMA and {\it NuSTAR}-FPMB into our fit: the final data set is composed of 33 spectra. 
When we fit the 33 spectra with the broad band model we get an overall $\chi^2$/dof=4378/3990=1.09 with no strong residuals in the 0.5-80 keV energy band (Figure \ref{reffit}). We find a best fit value for the black hole spin of $a$=$0.91^{+0.06}_{-0.07}$ and an inclination angle of the accretion disk $i$=$33\pm3 ^{\circ}$ (see Brenneman et al. in preparation, for a detailed analysis of the spin measurements from the same data set), in agreement with previous broad band analyses \citep{br06,miniutti07, chfa11}. We note that the spin errors are statistical only and do not include any systematics due to model degenaracies. The cross-calibration normalizations between the three detectors are $\rm K_{\rm Pn-FPMA}=1.081\pm0.007$ and $\rm K_{\rm Pn-FPMB}=1.112\pm0.006$. The best fit parameters are listed in Table \ref{refl_best_fit} (but note that a further analysis of interval 10 is presented in Sect. \ref{eclipses}). The addition of a high energy cutoff to the primary power law leads to an insignificant improvement in the fit and a lower limit of E$\rm _{C}>$110 keV is found. Precise measurements of the cutoff energy are treated in a separate work (Brenneman et al., in preparation).

  \begin{figure*}
\centering
\includegraphics[width=\columnwidth]{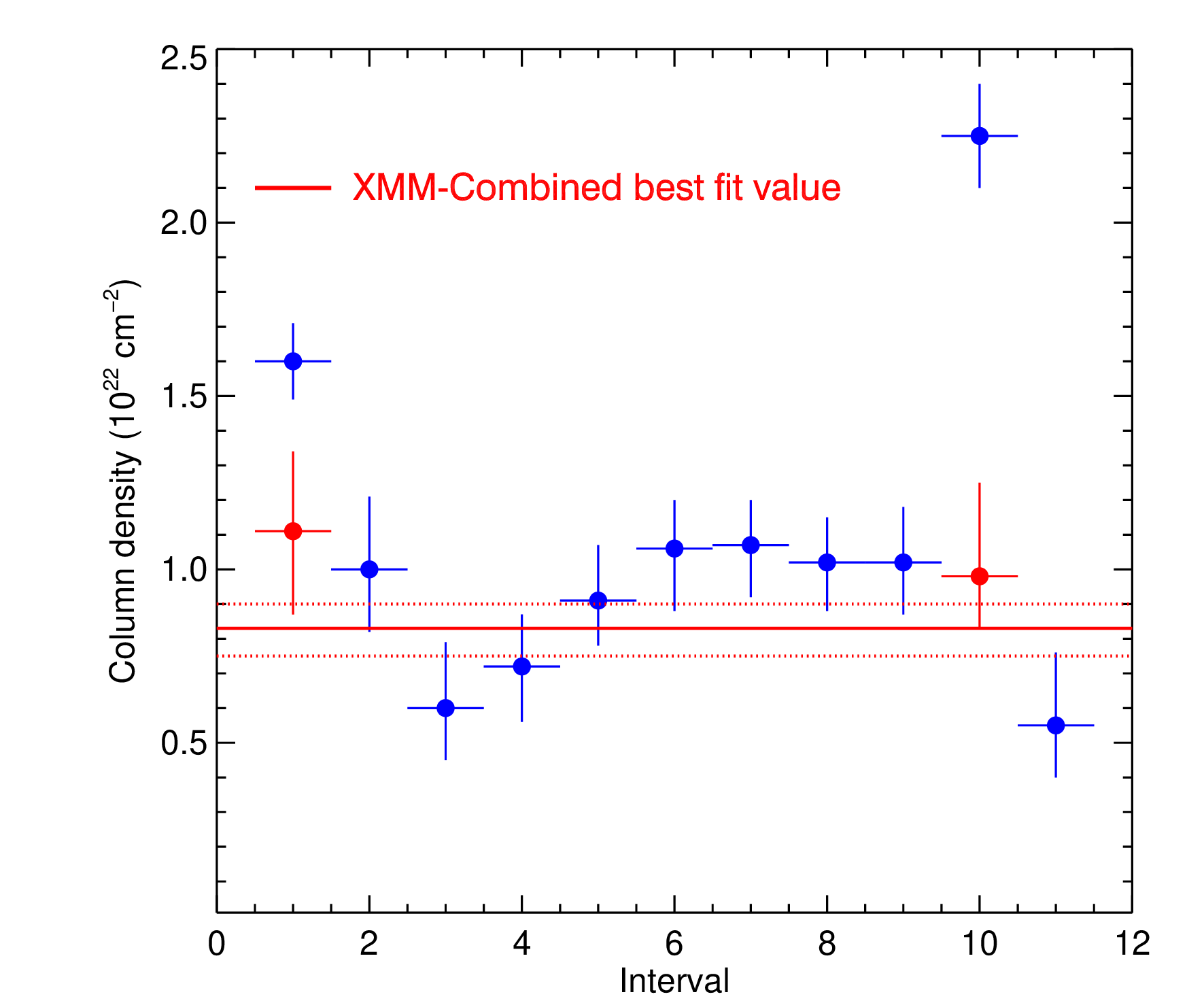}
\includegraphics[width=\columnwidth]{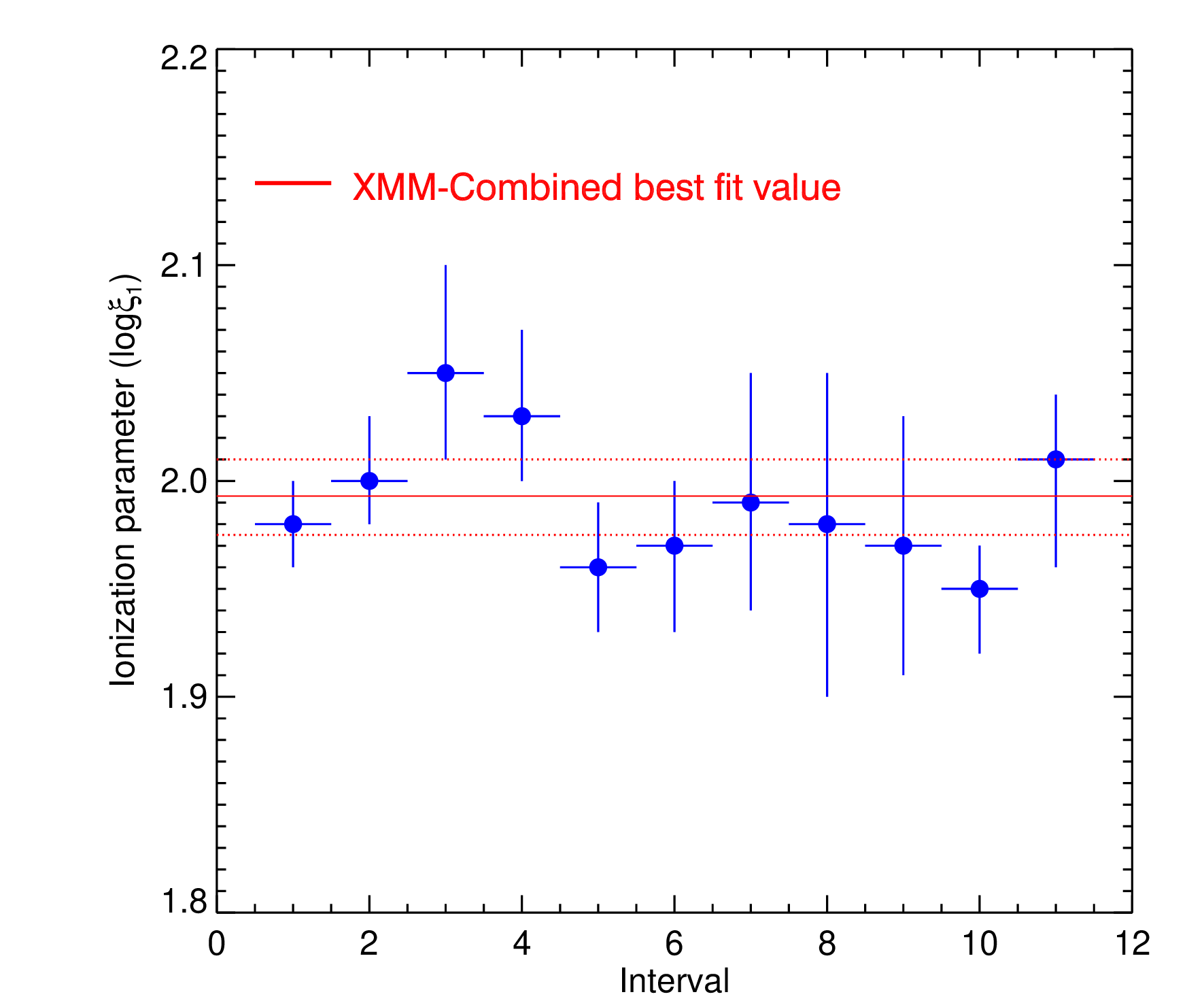}
\includegraphics[width=\columnwidth]{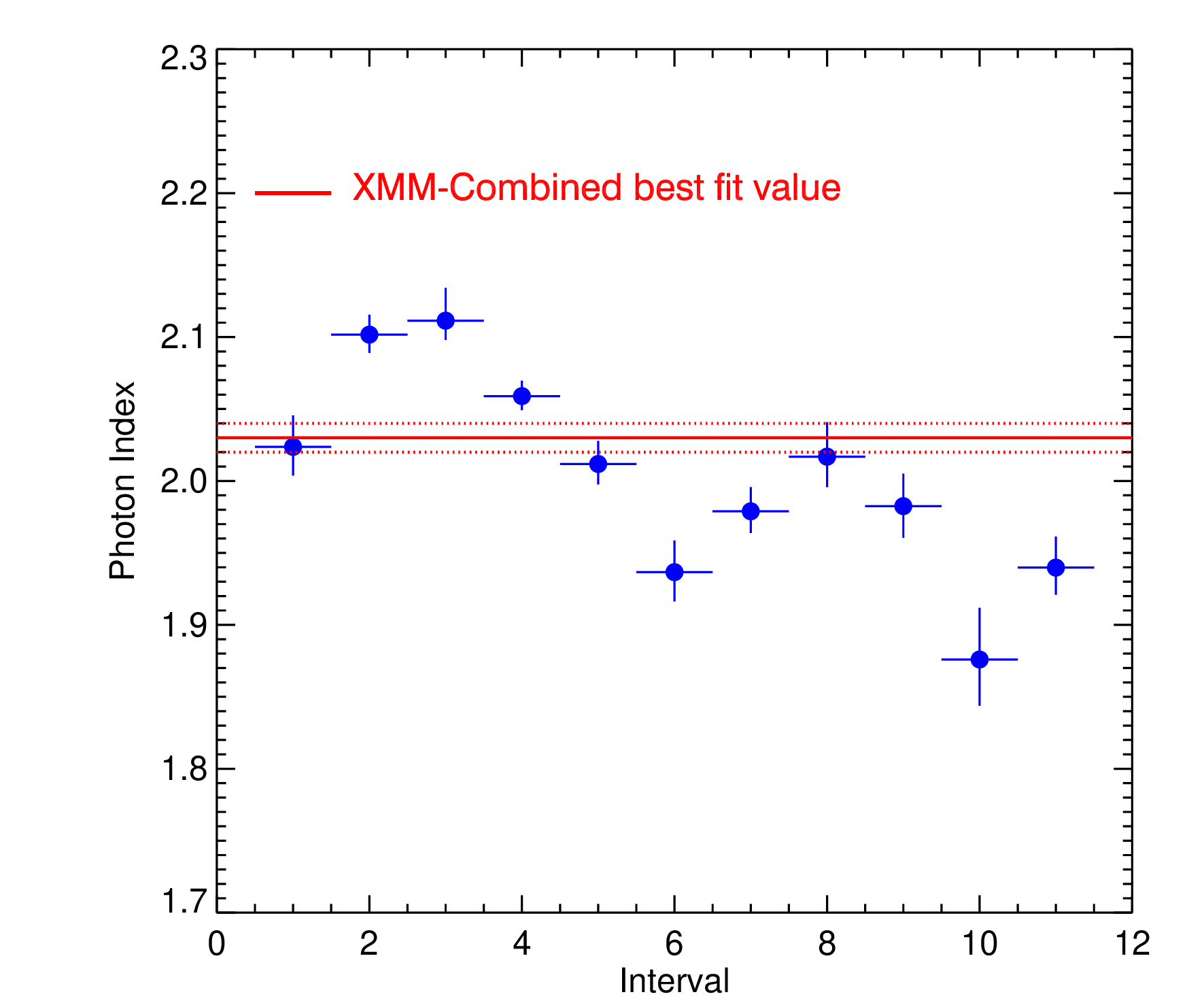}
\includegraphics[width=\columnwidth]{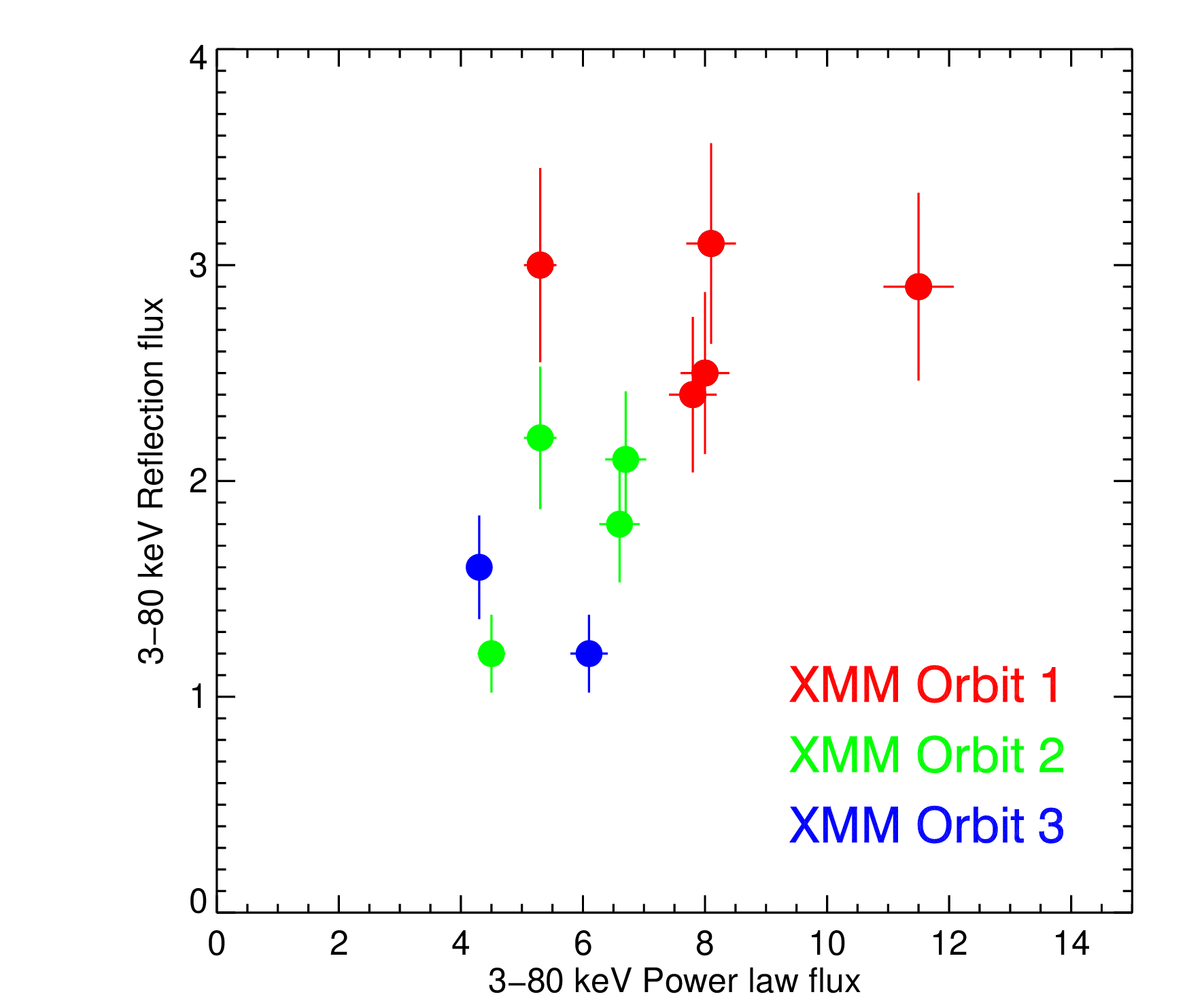}
\caption{\label{parameters_refl} \textit{Top panels:} light curves of the column densities (blue and red points indicate broad band best fit values and best fit values from the analysis described in Sect. \ref{eclipses}, respectively) and ionization parameters for the main warm absorber, throughout the observation. \textit{Bottom panels:} time evolution of the photon indices and reflection versus power law fluxes in 3-80 keV band, for the three {\it XMM} orbits, when a variable photon index of the primary continuum is considered. Fluxes, reported in Table \ref{fluxes}, are in $10^{-11}$ erg cm$^{-2}$ s$^{-1}$ units.}
\end{figure*}

Due to the large variation of the reflector's ionization state ($\log(\xi_{\rm refl}/ ({\rm erg}\ {\rm cm}\, {\rm s}^{-1}))=0.08-2.98$) we looked for a change of the photon indices between the 11 intervals. We tied the normalization of the cold reflector and the ionization parameters of the disk reflections between the 11 intervals and allowed the photon indices to vary. We get a best fit $\chi^2$/dof=4401/3999=1.10 and best fit values of $A_{\rm Fe} = 1.4 \pm 0.2$ and $\log(\xi_{\rm refl} / ({\rm erg}\, {\rm cm}\, {\rm s}^{-1})) = 2.9^{+0.1p}_{-0.2}$, where the $p$ indicates that the ionization parameter has pegged to the maximal value allowed in the model, $\log(\xi_{\rm refl} / ({\rm erg}\ {\rm cm}\, {\rm s}^{-1})) = 3.0$; we use this $p$ notation for the remainder of the text where model fits have pegged to limiting values in the models.

All others parameters are consistent with the ones presented in Table \ref{refl_best_fit}, within the errors. A total variation of the photon indices of $\Delta \Gamma\sim0.3$ is found (Fig \ref{parameters_refl}, bottom left panel) among the 11 intervals of the 2013 observation. This is not unexpected given the factor $\sim3$ change in the power law flux \citep{shemmer06,rye09}. Leaving the ionization parameters free does not improve the fit significantly. Fluxes for the reflection and primary components in the 3-80 keV energy band can be found in Table \ref{fluxes}.

We can estimate the black hole mass in MCG-6-30-15 from the variation in the slope of the continuum, using the $\Gamma$--$L_{\rm Bol}/L_{\rm Edd}$ relation in \citet{rye09}. Assuming a bolometric luminosity of $4\times10^{43}$ erg s$^{-1}$ \citep{rey97b} the estimated range of black hole masses is $2$--$7\times10^6$ M$_{\odot}$, in agreement with values in the literature \citep{mcha05,bennert06,ponti12}.

\begin{table}
\caption{\label{fluxes} Fluxes in $10^{-11}$ erg cm$^{-2}$ s$^{-1}$ units between 3 and 80 keV for the reflection and primary components.}
\begin{center}
\scriptsize
\begin{tabular}{ccc}
{\bfseries Interval } & $F^{\rm REF}_{\rm 3-80\ keV}$ & $F^{\rm PL}_{\rm 3-80\ keV}$  \\
 & & \\
\hline
\hline
1 &$3.0\pm0.4 $& $5.3\pm0.3$\\
2 & $2.5\pm0.4$& $8.0\pm0.4$\\
3 & $2.4\pm0.3$& $7.8\pm0.4$\\
4 & $2.9\pm0.4$& $11.5\pm0.6$\\
5 & $3.1\pm0.5$& $8.1\pm0.4$\\
6 &$2.2\pm0.3 $& $5.3\pm0.3$\\
7 & $1.8\pm0.3$& $6.6\pm0.3$\\
8 &$1.2\pm0.2$ & $4.5\pm0.2$\\
9 & $2.1\pm0.3$& $6.7\pm0.3$\\
10 &$1.6\pm0.2 $& $4.3\pm0.2$\\
11 & $1.2\pm0.2$& $6.1\pm0.3$\\
   \hline
  \hline
\end{tabular}
\end{center}
\end{table}

\subsection{Results}\label{reflres}
In the following section, we discuss the spectral variability of MCG-6-30-15 below and above 3 keV. The wide energy band (0.35-80 keV) available permits us, for the first time, to measure the parameters with high accuracy and to compare the behavior of the components in different energy intervals. 

The top panels of Figure \ref{parameters_refl} shows the best fit parameters for the main warm absorbing component. No variations in the ionization state of the material are present while in intervals 1 and 10 strong variations in the column density are present (blue points; the red points are explained in Sect. \ref{eclipses}). This variation in column density with a constant ionization parameter suggests the presence of neutral material partially eclipsing the soft X-rays ($<$3 keV) and leaving the hard part of the spectrum unchanged. If we look at the hardness ratio in Figure \ref{intervals}, interval 10 does indeed show a bell shaped structure with time suggesting an occultation event \citep[see][for the case of a cometary shaped event in NGC 1365]{mrs10}, possibly indicating a cloud crossing the line of sight on a time scale of $\sim$10-20 ks: this event is treated in greater detail in Sect. \ref{eclipses}. 

We now focus on the spectral variability above 3 keV, where the effects of the warm absorbers can be neglected. In the previous section we fitted the reflection model to the data set in two extreme assumptions: varying only the ionization state of the disk or the slope of the incident power law. In the former case most of the spectral variability can be attributed to changes in the ionization state of the accretion disk. We find only a trend where the ionization parameter $\xi_{\rm refl}$ is higher in the high flux states (Table 3). The changes are too large to be physical (three orders of magnitude in $\xi_{\rm refl}$ versus a variation of $\sim 3$ in the nuclear flux). A variation in the density profile of the disk on a time scale of hours is unlikely. 

On the other hand, when a variable slope of the continuum is taken into account, we find a modest change in $\Gamma$ between the different intervals (within $\Delta \Gamma\sim0.3$; Figure \ref{parameters_refl}, bottom left panel).  In Figure \ref{parameters_refl} (bottom right panel) fluxes in the 3--80 keV energy band of the reflection component are plotted against those of the primary power law (reported in Table \ref{fluxes}) for the three {\it XMM} orbits. During the first revolution (red data points), the flux of the reflection component is consistent with remaining constant, despite the variation of a factor $\sim$2 in the nuclear flux. This constancy of the reflection is in agreement with previous X-ray analyses \citep{vf04,miniutti07}, also in the case of variations in the nuclear continuum slope \citep{fv03,lfm07}. The behavior of the source in the first five time intervals can be explained in the framework of the light bending model \citep{mf04}, where the primary, variable emission is bent onto the accretion disk, which produces constant reflected emission. In the second and third revolutions (green and blue data points, respectively) variations in both the reflected and primary components are found, suggesting that the spectral variability of the source is intrinsic to the continuum X-ray emitter and not due to geometrical effects (i.e. the height of the X-ray source, or a varying spatial extent of the corona illuminating the disk). The light bending model, assuming non intrinsic flux variability, generally predicts a correlation at low fluxes and an almost constant reflection at higher fluxes. Our results show that this trend is observed, although the observed scatter implies that intrinsic flux variability is also present.

Interestingly, the change in normalization of the reflection components agrees with independent work based on a Principal Components Analysis of MCG-6-30-15 using multiple {\it XMM-Newton} observations \citep{parker13}. By analyzing 600 ks of total data the authors concluded that the observed relatively weak variability in the reflection component is, indeed, due to the effects of light-bending close to the event horizon of the black hole. Our analysis however uses only data from the 2013 {\it NuSTAR}+{\it XMM} campaign.

 Our parametrization is clearly an oversimplification: the density profile and a more complex ionization disk structure should be taken into account and connected to the geometrical properties of the primary X-ray emitter.  The variation of the height of the X-ray emitter in a lamp post geometry and its link with changes in the ionization state of the disk will be discussed in Brenneman et al. (in preparation). A complete analysis of the lags and reverberation properties of the source will then be presented in Kara et al. (in preparation).
 
 \begin{figure*}
\centering
\includegraphics[width=\textwidth]{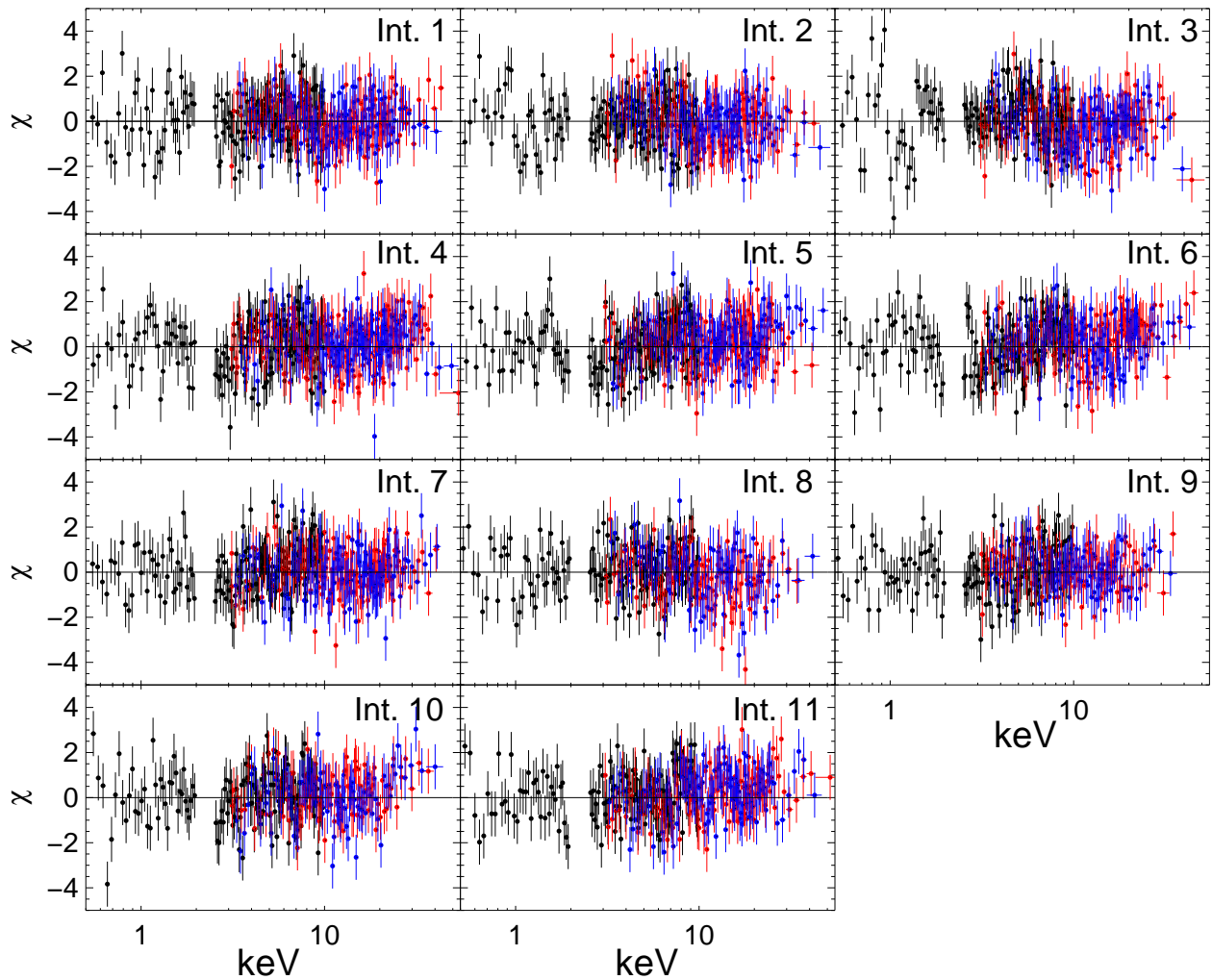}
\caption{\label{nufit} Residuals of the 33 {\it XMM} EPIC-Pn (in black) and {\it NuSTAR} FPMA (in red), FPMB (in blue) spectra with the absorption model.}
\end{figure*}
\begin{table*}
\caption{\label{abs_best_fit} Best fit parameters for the absorption model.\\ Column densities are in $10^{22}$ cm$^{-2}$ units, ionization parameters $\xi$ are in erg cm s$^{-1}$ units. In the top table joint best fit parameters are shown while in bottom table values for variable components are shown.
Columns:  
(a) Ionization parameter of the second fully covering warm absorber;
(b) Column density of the second fully covering warm absorber;
(c) Iron column density of the fully covering dusty absorber;
(d) Ionization parameter of the fourth warm absorber fully covering the distant ionized reflection component;
(e) Column density of the fourth warm absorber fully covering the distant reflection component ;
(f) Ionization parameter for the distant reflection component;
(g) Iron abundance with respect to the solar value;
(h) Normalization of the ionized reflection component;
(i) Photon index of the primary power law;
(l) Ionization parameter of the first fully covering warm absorber; 
(m) Column density of the first fully covering warm absorber;
(n) Ionization parameter of the fifth warm absorber partially covering the primary X-ray source;
(o) Variable column density of the fifth warm absorber partially covering the primary X-ray source;
(p) Normalization of the primary continuum component.
(q) Normalization of the absorbed primary component.
(r) Covering factor of the fourth warm absorber C$_{\rm F}=N_3/(N_2+N_3)$.}
\begin{center}
\scriptsize
\begin{tabular}{c|cc|c|cc|ccc|c}
   & & & & & & & & \\
 & $\log (\xi_2)$ &$ {\rm N_{H_2}}$& $\log ({\rm N_{Fe}})$&$\log (\xi_4)$& $ {\rm N_{H_4}}$&$\log ({ \xi_{\rm REFL }})$& ${\rm A_{Fe}}$&N$_1$ ($\times10^{-6}$)&$\Gamma$\\
 {\bfseries Interval }& (a) & (b) & (c) & (d) & (e) & (f) & (g) & (h) &(i)   \\
 \hline
 \hline
   & & & & & & & & \\
 1-11&$1.47^{+0.03}_{-0.04}$ & $0.10^{+0.04}_{-0.04}$& $17.17^{+0.08}_{-0.09}$&$2.16_{-0.01}^{+0.01}$ &$38.1^{+1.6}_{-1.2}$ &$2.35^{+0.05}_{-0.05}$&$0.5^{+0.04}_{p}$ &$1.3^{+0.1}_{-0.1}$& $2.155_{-0.005}^{+0.007}$\\
 & & & & & & & & \\
\hline
\hline
\end{tabular}\\
$\ \ \ \ \ \ \ $\\
$\ \ \ \ \ \ \ $\\
$\ \ \ \ \ \ \ $\\
\begin{tabular}{c|cc|ccccc}
 & $\log (\xi_1)$& ${\rm N_{H_1}}$ &$\log (\xi_5)$& ${\rm N_{H_5}}$&N$_2$ $(\times10^{-2})$&N$_3$ $(\times10^{-2})$& C$_{\rm F}$\\
 {\bfseries Interval } &(l)& (m) & (n) &(o)&(p)&(q)&(r) \\
 \hline
 \hline
   & & & &  & \\
    1&$1.99^{+0.02}_{-0.01}$ & $1.03^{+0.03}_{-0.03}$ &$0.60_{-0.21}^{+0.12}$ & $15.9^{+18.6}_{-14.7}$&$1.59^{+0.04}_{-0.04}$ &$0.20^{+0.04}_{-0.04}$ & 11\%\\
     & & &  & & \\
     2& $2.02^{+0.02}_{-0.02}$& $1.34^{+0.06}_{-0.06}$&$0.60^*$ & $<1.5$ &$2.48^{+0.03}_{-0.03}$ & $<0.035$&$<1.4$\% \\
         & & & & & & \\
     3&$2.01^{+0.02}_{-0.02}$ & $1.35^{+0.07}_{-0.07}$& $0.60^*$& $<1.5$& $2.41^{+0.03}_{-0.03}$&$<0.045$ & $<1.8$\%\\
         & &  & & & \\
     4& $2.00^{+0.02}_{-0.02}$&$0.81^{+0.09}_{-0.07}$ & $0.60^*$&$1.7^{+0.5}_{-0.3}$ &$2.83^{+0.02}_{-0.03}$ & $0.65^{+0.05}_{-0.05}$&19\% \\
         &  & & & & \\
     5&  $1.96^{+0.02}_{-0.01}$&$0.73^{+0.12}_{-0.09}$ & $0.60^*$& $2.2^{+0.5}_{-0.6}$& $2.02^{+0.03}_{-0.04}$& $0.51^{+0.05}_{-0.05}$&20\% \\
         &  & & & & \\
     6& $1.91^{+0.05}_{-0.05}$&  $0.45^{+0.11}_{-0.09}$& $0.60^*$&$2.8^{+0.4}_{-0.4}$ & $1.03^{+0.02}_{-0.02}$& $0.45^{+0.03}_{-0.03}$&30\% \\
         & &  & & & \\
     7& $2.01^{+0.03}_{-0.03}$& $0.55^{+0.11}_{-0.10}$&  $0.60^*$&$2.1^{+0.3}_{-0.3}$ & $1.24^{+0.03}_{-0.03}$& $0.55^{+0.03}_{-0.03}$& 31\%\\
         & & &  & & \\
     8& $1.98^{+0.04}_{-0.04}$&$0.68^{+0.18}_{-0.13}$& $0.60^*$ & $1.9^{+0.6}_{-0.6}$&$1.01^{+0.03}_{-0.03}$ & $0.20^{+0.03}_{-0.03}$&17\% \\
         & &  & & & \\
     9& $1.98^{+0.04}_{-0.04}$& $0.78^{+0.13}_{-0.11}$ & $0.60^*$&$3.6^{+0.9}_{-0.8}$ & $1.41^{+0.04}_{-0.04}$& $0.47^{+0.05}_{-0.05}$& 25\%\\
 & & &  & & \\
      10 & $1.88^{+0.11}_{-0.09}$&$0.31^{+0.22}_{-0.14}$ &$0.60^*$ & $4.4^{+0.6}_{-0.5}$&$0.50^{+0.03}_{-0.03}$ & $0.53^{+0.03}_{-0.03}$& 51\%\\
 & &  & & & \\
      11 & $1.98^{+0.05}_{-0.06}$&$0.52^{+0.12}_{-0.13}$ & $0.60^*$&$1.8^{+0.3}_{-0.3}$ & $0.98^{+0.03}_{-0.03}$& $0.51^{+0.03}_{-0.03}$& 34\%\\
 & &  & & & \\
\hline
\hline
\end{tabular}\\

\end{center}
\end{table*}
 
\section{Testing the absorption scenario}
An alternative interpretation of the complex spectral variability of MCG-6-30-15 has been given by \citet{mtr08, mtr09} in terms of complex absorbing structures along the line of sight. In this section we discuss the application of this model to the joint {\it NuSTAR}-{\it XMM} data set. The complex absorption features below 3 keV have been fitted with two warm absorbers and a dusty absorber, fully covering the nuclear X-ray source: these components are the same used in Sect. \ref{rgs} to fit the higher resolution spectra from the RGS. The red wing of the Fe K$\alpha$ line and the strong Compton hump must then be interpreted in terms of additional absorbing regions. We introduce a fourth absorber in our model that fully covers the distant, ionized material responsible for the emission of the Fe K$\alpha$ line. The last absorber partially covers the nuclear X-ray source and, in this physical scenario, responsible for the spectral variations. The \textsc{xstar} tables used to reproduce the warm absorbers are the same ones described in Sect. \ref{analysis}. 

In \textsc{xspec} the model reads as follows: \\\\
\textsc{tbabs}$\times$\textsc{warmabs$_1$}$\times$\textsc{warmabs}$_2\times$ \textsc{dustyabs}$\times$\\
(\textsc{warmabs$_4$}$\times$\textsc{xillver}+\textsc{warmabs$_5$}$\times$\textsc{zpow}+ \textsc{zpow})\\\\
and is shown in Figure \ref{models} (right panel).
The covering factor of the fifth absorber is calculated as ${\rm C_F=N_{ABS}/(N_{ABS}+N_{UNABS})}$ where ${\rm N_{ABS}}$ is the normalization of the absorbed power law and ${\rm N_{UNABS}}$ is the normalization of the unabsorbed nuclear component.\\
  \begin{figure*}
\centering
\includegraphics[width=\columnwidth]{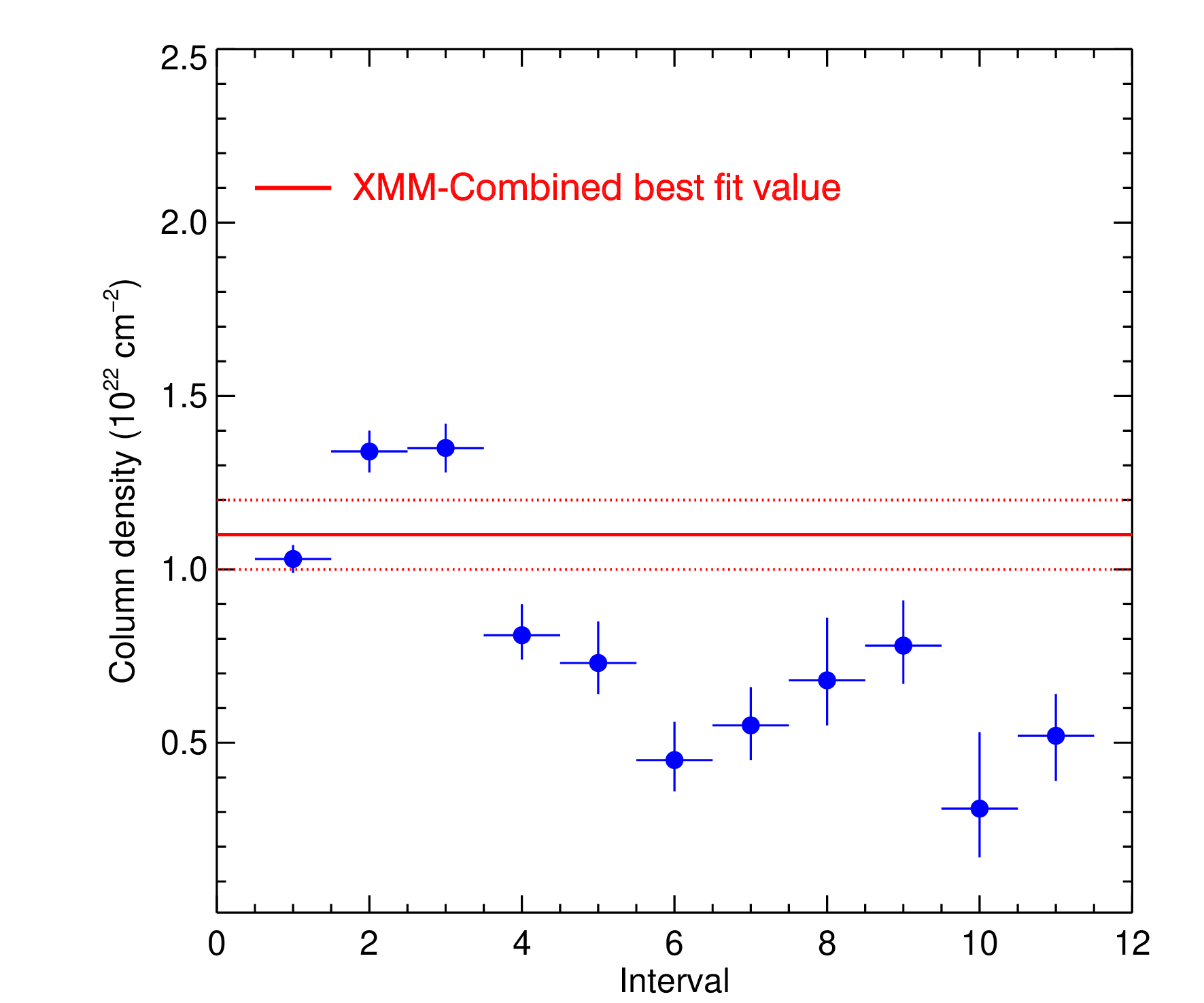}
\includegraphics[width=\columnwidth]{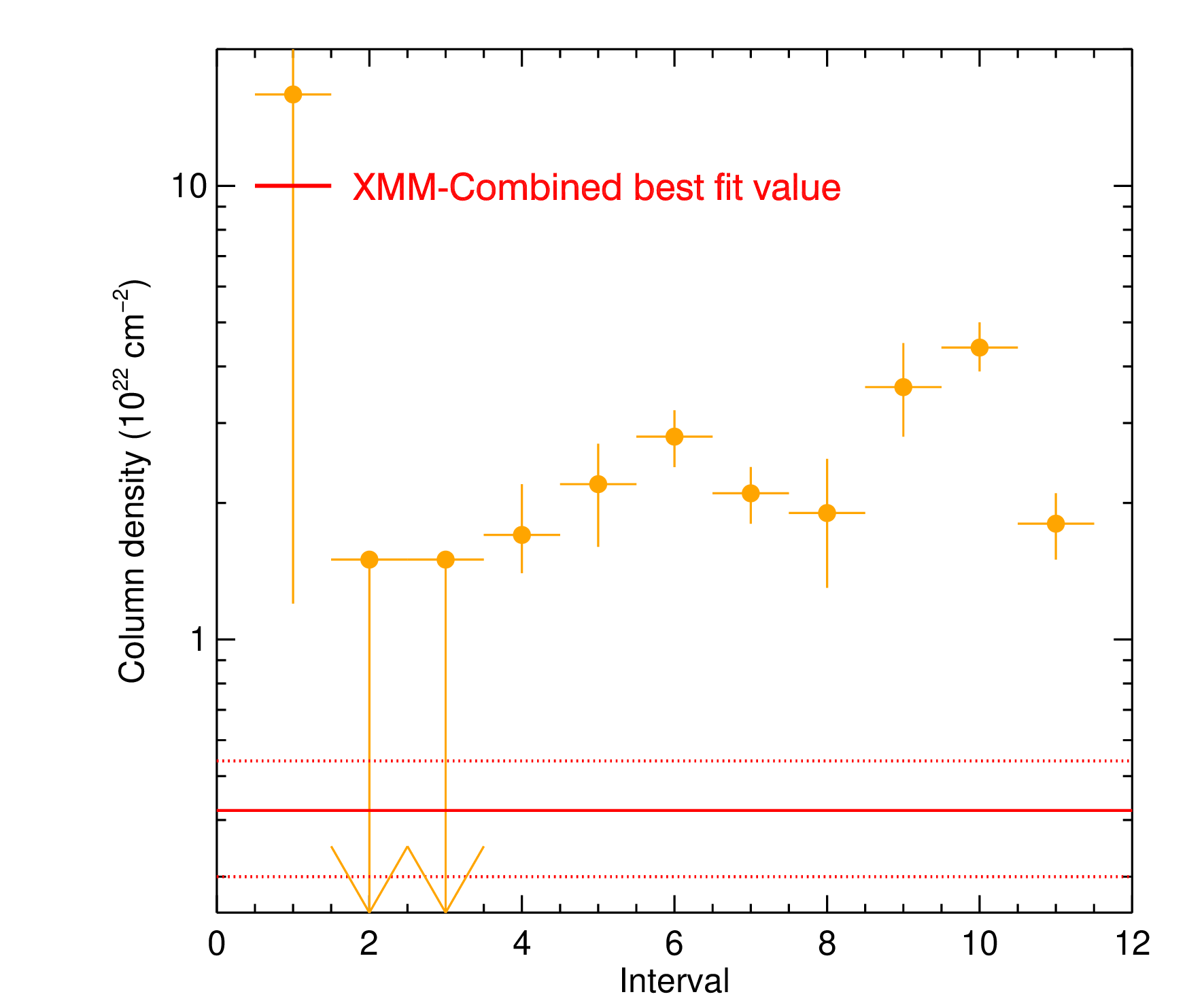}
\includegraphics[width=\columnwidth]{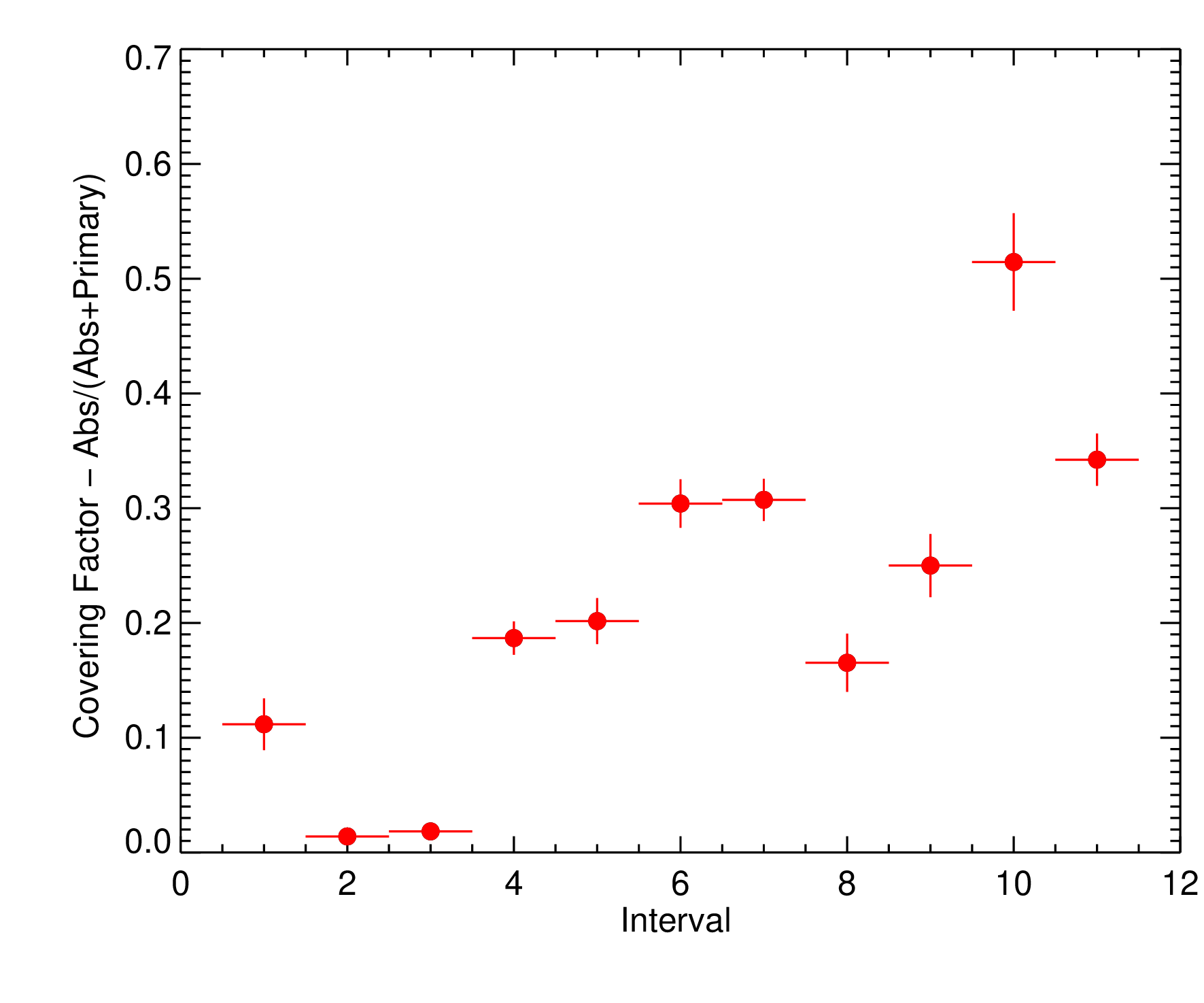}\\
\includegraphics[width=\columnwidth]{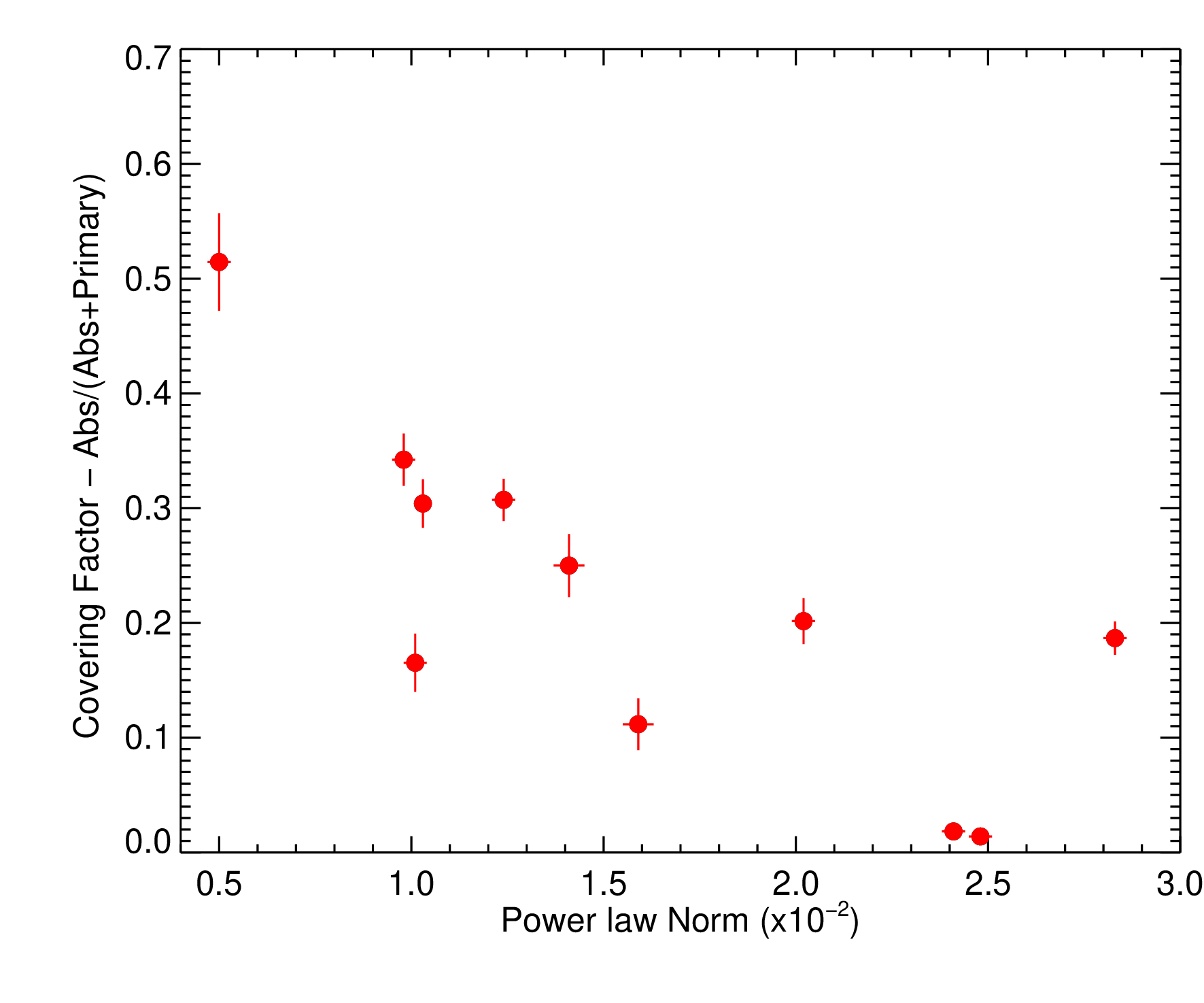}
\includegraphics[width=\columnwidth]{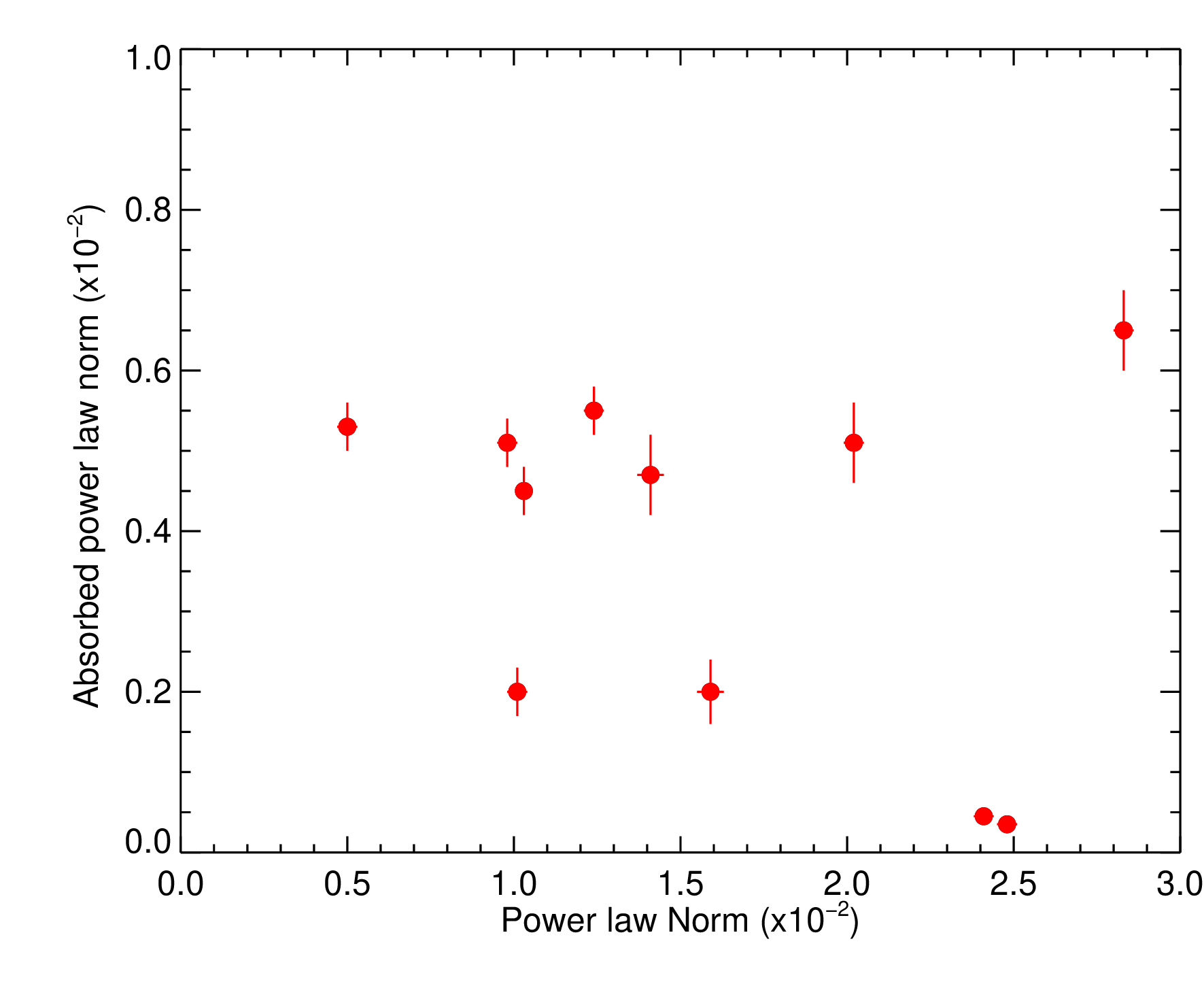}
\caption{\label{parameters_abs} \textit{Top panels:} light curves of the column densities of the fully covering warm absorber ($\rm N_{H1}$, left panel) and of the partial covering warm absorber ($\rm N_{H5}$, right panel). \textit{Middle panel:}  time evolution of the covering factor C$_{\rm F}$. \textit{Bottom panels:} The covering factor versus the normalization of the primary power law is plotted in the left panel while amplitudes of the absorbed and primary components are plotted in the right panel. Horizontal lines in top panels indicate best fit values (solid lines) and uncertainties (dashed lines) from the {\it XMM-Newton} best fits.}
\end{figure*}

\subsection{Broadband spectral anaslysis}
We first fit the 11 EPIC-Pn spectra leaving the parameters of the warm absorbers free to vary and the normalizations of the absorbed and primary power laws as the only variable. In this way we attribute all the spectral variations to the partial covering of the primary continuum. We get $\chi^2$/dof=2731/1605=1.70, and systematic residuals are present throughout the energy band. We get a best fit value for column density of the partial covering absorber of $\rm{N_H{_5}}= 4.2\pm1.2 \times 10^{21}$ cm$^{-2}$, ten times smaller than the one found in \citet{mtr08}. A simple model where partial covering of the X-ray source is the only variable component between the 11 intervals is not enough to reproduce the spectral complexity of the source: we therefore leave the column density of the partial covering absorber ($\rm{N_H{_5}}$) and of the most intense warm absorber ($\rm{N_H{_1}}$) free to vary. The fit improves ($\Delta\chi^2=652$) and the residuals left in the spectra are at energies smaller than 2 keV: the best fit  $\chi^2$/dof=2079/1577=1.31 (see Figure \ref{xmmfit}, right panel). We then introduced the 22 {\it NuSTAR} FPMA and FPMB spectra in the fit and we get a slightly worse overall fit ($\chi^2$/dof=4610/4019=1.15) with respect to the reflection model  (Figure \ref{nufit}). The cross-calibration factors between the three detectors are $\rm K_{\rm Pn-FPMA}=1.084\pm0.007$ and $\rm K_{\rm Pn-FPMB}=1.114\pm0.007$, consistent with the values found in Sect. \ref{fulldataset}. Best fit parameters can be found in Table \ref{abs_best_fit}.
 Since the fourth absorber, fully covering the ionized reflected emission, is the one with the highest column density ($3.8\pm0.1\times 10^{23}$ cm$^{-2}$) we let this parameter free to vary but the improvement is marginal with respect to the best-fit.

\subsection{Results}
Table \ref{abs_best_fit} shows broad band best fit parameters for the absorption model. In the context of this physical scenario we find that the spectral variability of MCG-6-30-15 does not arise exclusively from variations in the circumnuclear material partially covering the X-ray emitting source. 

 When we compare our best fit parameters with the ones in \citet{mtr08} some differences are found.We measure a flatter photon index (a $\Gamma=2.265\pm0.017$ was reported in the past {\it Suzaku} data) and we do not observe a third fully covering, highly ionized warm absorber, since no iron XXV K$\alpha$ and iron XXVI K$\alpha$ absorption lines are detected.

The top left and right panels of Figure \ref{parameters_abs} show the column density of the fully covering warm absorber and of the partial covering warm absorber (parameters $\rm N_{H1}$ and $\rm N_{H5}$ in Table \ref{abs_best_fit}), respectively. Variation in both absorbers can be seen and since there is a clear interplay between the two components it is not possible to draw any conclusion about their physical distance from the nucleus with any reasonable precision. 

One way to roughly parametrize the distance is to consider the time scales of the variations. The light travel time for one gravitational radius ($R_G=GM/c^2$) is $t=R_G/c\simeq 23 s$ (assuming a black hole mass of 5$\times 10^6$ M$_{\odot}$). We can calculate a lower limit for the distance of the emitting region if we consider the two closest time intervals with the greatest variation. For instance, between intervals three and four a $\Delta {\rm N_{H1}}\simeq0.5\times10^{21}$ cm$^{-2}$ is found and only an upper limit is found for $\rm N_{H5}^{\rm Int. 3}$. The difference between the two intervals is $\sim 30$ks which is equivalent to a lower limit of $\sim1300 \ {\rm R_G}$ to the variability length scale. This value is consistent with the time scale we investigate in Sect. \ref{eclipses} but we stress the fact that not all the spectral variability is due to occultation effects. 

 The main difference between our work and previous analyses \citep{mtr08} is the lack of coherence between the variation of the amplitude in the partially absorbed component and the direct continuum. If we plot the normalizations of the two components (Figure \ref{parameters_abs}, bottom right panel) we do not see the linear trend observed in the past (note that the y-axis and x-axis have different scales, for the sake of clarity). We inferred a C$_{\rm F}$ in the 0-50\% range while it was suggested to be between 50-100\% in the previous broad band analysis. We find no coherent variation between C$_{\rm F}$ and the amplitude of the direct nuclear component (Figure \ref{parameters_abs}, bottom left and middle panels). This different behavior can be attributed to the high flux state in which we observed the source and the complex interplay between the parameters. The way we calculate the covering factor (${\rm C_F=N_{ABS}/(N_{ABS}+N_{UNABS})}$) is clearly an oversimplification and this particular parameter is very model dependent. It depends strongly on the flux state of the source and the circumnuclear geometrical structure: the sizes of the X-ray emitter and of the absorber, together with its ionization state. We therefore conclude that, in the absorption scenario, the X-ray spectral variability of MCG-6-30-15 is not due only to variable partial covering but it is also related to the complex interplay between the intrinsic variable continuum and the circumnuclear (still unknown) geometry on scales greater than $\sim1300 \ {\rm R_G}$.

 The absorption model for MCG--6-30-15 has been slightly revised in \citet{mtr09}, where the ionized reflected emission has been removed and replaced by a layer of neutral absorbing material, partially covering the nuclear emission. We tested this by modifying the absorption model used in previous sections. We replaced the \textsc{xillver} component of our model (the one reproducing the ionized reflection) with a Gaussian line for the Iron K$\alpha$ emission line, fixing its width to the value measured with {\it Chandra} HEG \citep{yaq04}. The fourth absorber is now considered as partially covering the nuclear continuum. The model is hence composed by five layers of absorbing material: three of them have covering fraction $c_f=1$ (the two warm absorbers and the dusty one), responsible for the absorption lines below ~3 keV and two of them (the ones responsible for the apparent broadening of the iron K$\alpha$ line) are partially covering the nuclear emission. In this scenario the column densities and covering factors of the fourth and fifth absorbers, together with the normalizations of the power law are left free between the 11 intervals, while the other parameters were tied. We find a best fit $\chi^2$/dof=5270/4022=1.31. The fit is poor mainly due to residuals below 3 keV, suggesting that there is a complex interplay between the innermost and outermost layers of absorbing material.

\section{Broad Line Region eclipses}\label{eclipses}
Absorption variability is often found when we compare observations months to years apart \citep{risa02b}, and, most notably, has been found on time scales of hours to days in several sources, such as NGC1365 \citep{ris05, ris07, ris09}, NGC 4388 \citep{elvis04}, NGC~4151 \citep{puc07} and NGC 7582 \citep{bianchi09c}. Very recently a  homogeneous analysis of a statistically representative sample of AGN has been carried out in Risaliti~et al. (in preparation) and eclipses from Broad Line Region (BLR) clouds have been found in a number of sources (MCG-6-30-15, NGC 3783 and NGC3227 among  others). The presence of BLR eclipsing material has also been found in SWIFT J2127.4+5654, a bright Sy 1 galaxy well-known for its broad Fe K$\alpha$ line \citep{sanmi13}.

In the following, we discuss some of the extreme variations in the hardness ratio of MCG-6-30-15 in terms of a cloud crossing the line of sight. It is beyond the scope of this analysis to look for systematic occultations throughout our observation but we use the variation of the column density of the warm absorber in Figure \ref{parameters_refl} (top left panel) as evidence of a possible eclipse.

\begin{table}
\caption{\label{blrfit} Best fit parameters with X-ray occultation.\\ Column densities are in $10^{22}$ cm$^{-2}$ units, ionization parameters $\xi$ are in erg cm s$^{-1}$ units.}
\begin{center}
\scriptsize
\begin{tabular}{cccc}
{\bfseries Parameter } & Int. 1a & Int. 6a&Int. 10  \\
\hline
\hline
 N$_{\rm H1}$ & $1.11^{+0.23}_{-0.24}$&$0.53^{+0.18}_{-0.21}$ &$0.98^{+0.27}_{-0.15}$ \\
 $\log(\xi_1)$&$1.97^{+0.03}_{-0.03}$ & $2.06^{+0.12}_{-0.12}$&$1.91^{+0.04}_{-0.04}$ \\
 N$_{\rm H_{cloud}}$&$<0.05$& $<0.05$&$2.2^{+0.8}_{-0.5}$ \\
C$\rm _F$& $<0.1$&$<0.1$ &$0.32^{+0.03}_{-0.05}$ \\
$\log ({ \xi_{\rm refl }})$ & $2.75^{+0.15}_{-0.27}$& $1.04^{+0.15}_{-0.21}$&$1.01^{+0.21}_{-0.26}$ \\
 N$_{\rm refl }$ ($\times 10^{-5}$)& $0.06^{+0.02}_{-0.04}$&$2.0^{+0.3}_{-0.3}$ &$3.8^{+0.3}_{-0.1}$\\
 N $_{\rm POW }$($\times 10^{-2}$) &$1.9\pm0.02$&$0.91\pm0.02$ &$0.79\pm0.03$\\
   \hline
  \hline
\end{tabular}
\end{center}
\end{table}

\begin{figure}[h!] 
\centering
\includegraphics[width=0.7\columnwidth, angle=-90]{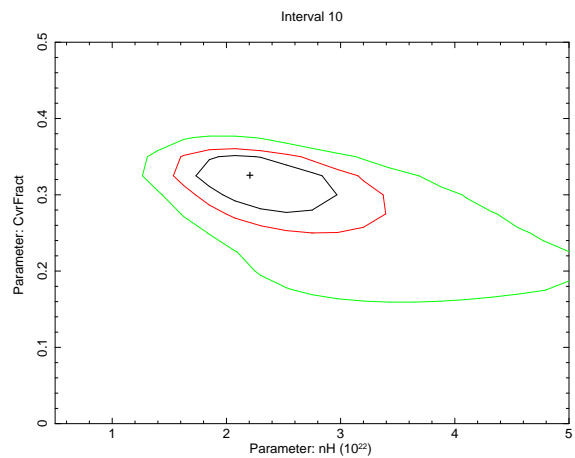}
\caption{\label{contours} Contour plots of covering factor versus column density in the spectrum extracted from interval 10.  Solid black, red and green lines corresponds to 68\%, 90\% and 99\% confidence levels, respectively.}
\end{figure}

Interval 10 is the one where the measured ${\rm N_{H1}}$ is largest ($2.25\pm0.15\times10^{22}$ cm$^{-2}$) and where the hardness ratio reaches its maximal value ($\sim0.52$). For this interval, the best fit with a reflection model leads to a $\chi^2$/dof=370/318=1.16 (parameters can be found in Table \ref{refl_best_fit}). We model the new absorbing component with the \textsc{zpcfabs} component in \textsc{XSPEC}, leaving the column density and covering factor as free parameters. 
Additional variable components are the power law normalization, the ionization parameter and the column density of the principal warm absorber, the ionization parameter of the reflection ($\log ({ \xi_{\rm refl }})$) and its normalization. Other parameters are fixed to their best fit values. The best fit $\chi^2$/dof is 340/319=1.06 and the best fit parameters are presented in Table \ref{blrfit}. The best fit parameter for the column density of the warm absorber is now consistent with the {\it XMM}-combined one (Figure \ref{parameters_refl}, top left panel, red data point). The contour plot between the column density and the covering factor of the cloud can be seen in Figure \ref{contours}. There is also a marginal variation in the ionization parameter of the reflection. It is worth comparing this particular time interval with another two: the first one is interval 1, where a change of column density is found, and interval 6a (Figure \ref{intervals}) where the source is in a similar flux state.
 \begin{figure}[h!] 
\centering
\includegraphics[width=\columnwidth]{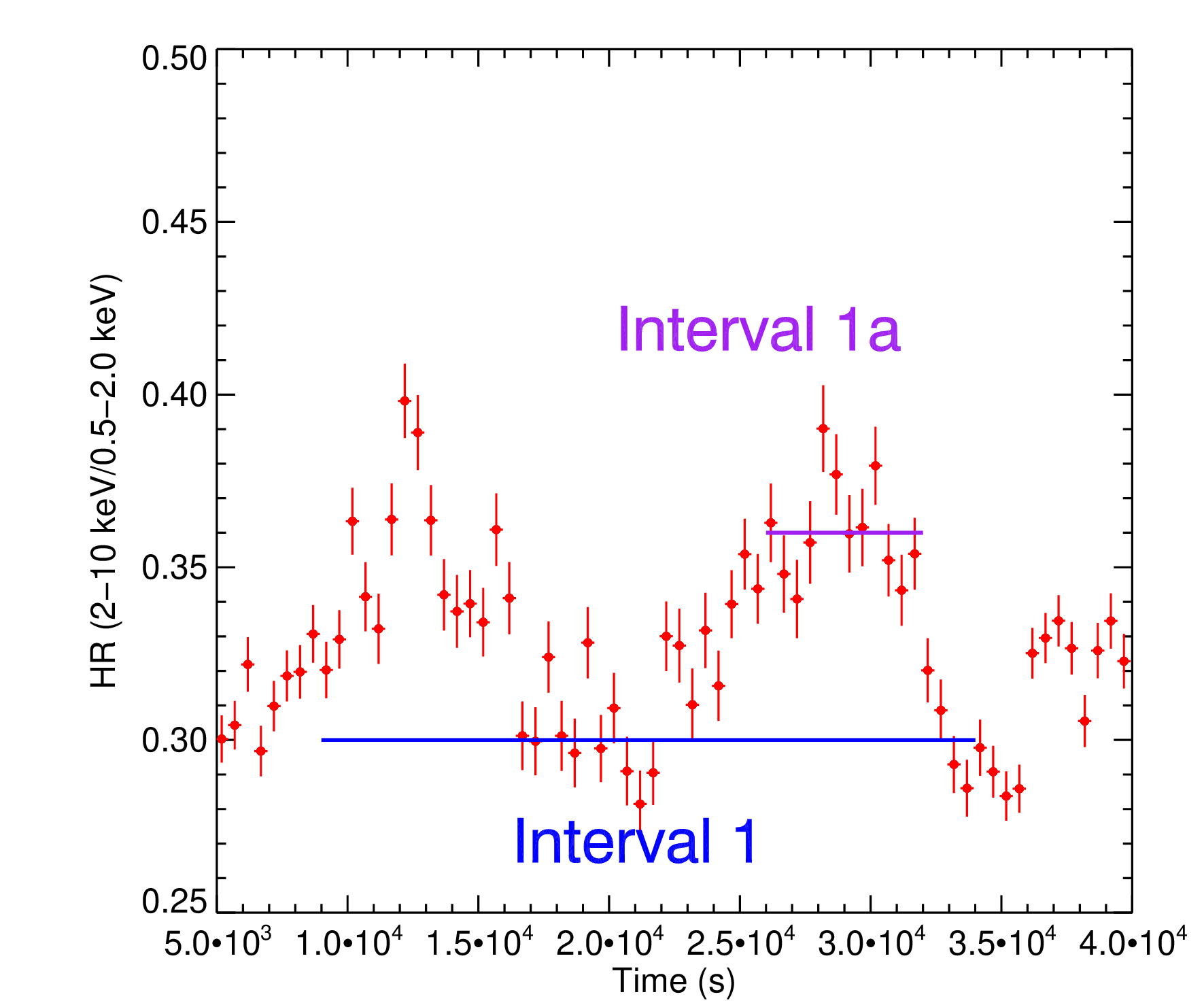}
\caption{\label{hr1} Zoomed hardness ratio between 500 s and 4000 s from the start of the observation. The change in spectral shape cannot be attributed to a BLR cloud partial covering the line of sight but to the complex interplay between the primary and disk reflection components.}
\end{figure}
 \begin{figure}[h!] 
\centering
\includegraphics[width=0.7\columnwidth, angle=-90]{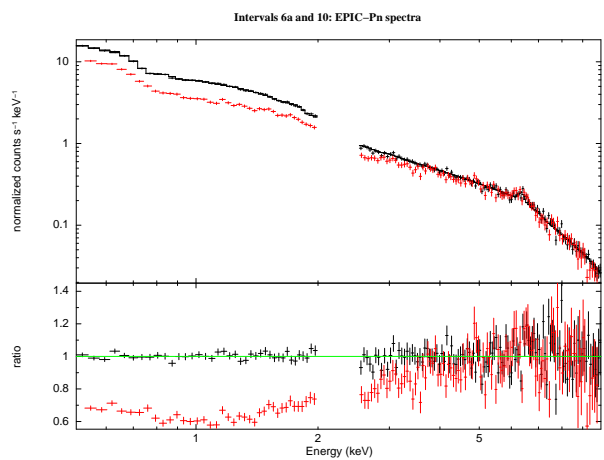}
\caption{\label{eclipse} Spectra from intervals 6a (in black) and 10 (in red) are shown. In the lower panel the ratio between the model used to fit interval 6a and the spectrum extracted from interval 10 is shown. The shape of an X-ray occultation can be seen. }
\end{figure}

During the first interval of our analysis a column density of ${\rm N_{H1}}$=$1.6\pm0.1\times10^{22}$ cm$^{-2}$ is found. The hardness ratio is plotted in Figure \ref{hr1}. We extracted an EPIC-Pn spectrum from interval 1a and fit it with the above model. The best fit $\chi^2$/dof is 162/133=1.2 and only upper limits for the eclipse are found. Free parameters can be found in Table \ref{blrfit} and indicate a different physical origin for the change in hardness ratio. This is not due to an occultation event from BLR clouds but to relative changes in the amplitudes of the primary and reflection components with respect to the best fit values found for interval 1.

When we apply the model of the X-ray occultation to the EPIC-Pn spectrum extracted from interval 6a  we find a good fit, $\chi^2$/dof=168/141=1.19, and only upper limits for the density and covering factor of the clouds (Table \ref{blrfit}). In Figure \ref{eclipse} the spectra of interval 6a and 10 are plotted and the typical spectral effect of the eclipse may be seen.

Information on the physical properties of the eclipsing cloud can be derived from the occultation observed in interval 10. Following the kinematic considerations extensively described in \citet{ris07} and Risaliti et al. (in preparation), assuming that the nuclear X-ray source has a linear size $D_S=5 R\rm _G$ \citep{rm13}, the transverse velocity of the cloud can be calculated as $v\sim 1.5\times 10^3\ M_6\ t_4^{-1}\ (\Delta C_F)^{1/2}$ km s$^{-1}$, where $M_6$ is the black hole mass in 10$^6$ M$_{\odot}$ units, t$_4$ is the occultation time in 10 ks units and $\Delta C_F$ is the covering factor variation during the occultation. If we use $M_6=5$ (see Sect. \ref{fulldataset}), an occultation time of $\sim$20 ks (elapsed duration of interval 10) and $\Delta C_F=0.32$, we infer a transverse velocity $v\simeq 3\times 10^3$ km s$^{-1}$. If we then consider the absorbing material located at a distance $R$ from the central X-ray source, moving with Keplerian velocity ($v=v_K$), we can calculate $R=GM_{\rm BH}v_K^{-2} \simeq7\times 10^{15}$cm$\simeq10^4\ R_G$. The cloud density is then $n\sim N_H/D_S\simeq 7\times10^{9}$ cm$^{-3}$. These estimates of velocity, distance from the X-ray source and density are consistent with values typically inferred for BLR clouds and agree with the analysis performed in Risaliti et al. (in preparation). 

Occultation by BLR clouds does not change the conclusions of the reflection scenario described in Sect. \ref{reflres}, where the eclipses are taken into account with a change of the column density of one of the two ionized absorbers. It is also worth noting that the presence of BLR clouds along the line of sight does not interfere with measurements of the properties of the black hole spin, as discussed for the case of NGC1365 in \citet{risa13} and Walton et al. (submitted), but may be used to study the broad iron K$\alpha$ line \citep{rne11} .

\section{Flux-flux plots}\label{diffspectra}
We calculate flux-flux plots \citep[see][]{tum03} to compare the count rate in the \emph{NuSTAR} and \emph{XMM-Newton} energy bands. We compare two bands from each instrument: 3--10 keV and 10--50 keV from \emph{NuSTAR} with 0.5--2 keV and 2--10 keV from \emph{XMM-Newton}, for each of the 11 intervals. The fluxes and errors are calculated using \textsc{cflux} in \textsc{Xspec}.

We find that all four flux-flux plots are well fit with a simple linear relationship between the fluxes in different bands, although with significant scatter, with no improvement in the reduced $\chi^2$ by fitting a powerlaw model. Curvature in the flux-flux plots would be indicative of pivoting of the spectrum with flux \citep[as found by][for NGC 4051]{tum03}.
In all four plots the best fit line is offset from zero by a positive amount. This indicates the presence of a relatively constant hard component, which remains in the {\it NuSTAR} bands after the main variable component is subtracted.

\begin{figure}
\centering
\includegraphics[width=\columnwidth]{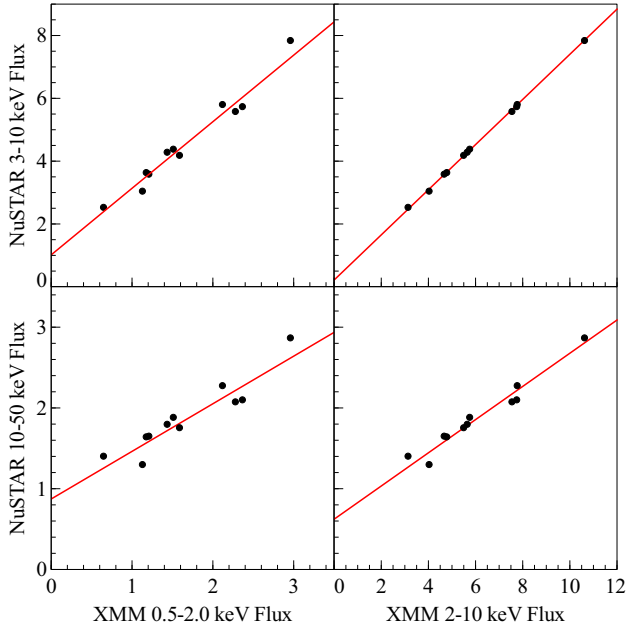}
\caption{\label{fluxplots}Count rate in two \emph{NuSTAR} energy bands (3--10 keV, top, and 10--50 keV, bottom) plotted against the count rate in two \emph{XMM-Newton} energy bands (0.5--2 keV, left, and 2--10 keV, right). The red lines show the best linear fits to the data. Error bars are not shown, as they are smaller than the points, but are on the order of 1~per cent for the lowest flux interval in the 10--50 keV band, and smaller for the other points.}
\end{figure}

The scatter is very small in the plot comparing the overlapping 2--10 and 3--10 \emph{XMM} and \emph{NuSTAR} bands, and is too large in the other plots to be consistent with noise. This implies that there is some spectral variability that is largely uncorrelated with flux and which affects the hard and soft bands differently. To investigate the origin of this scatter we calculate the same figures using spectra extracted using 200~s intervals as in \citet{vf04}, then binned by flux, to remove the effects of variability uncorrelated with flux. Using this method, we find no significant deviation from the linear fits and conclude that the scatter in Fig.~\ref{fluxplots} is due to flux independent variability.

Such variability could be caused by several mechanisms which do not affect the soft and hard bands equally, including absorption or reflection variability, or pivoting of the power law continuum. We can examine the nature of this variability by calculating a \emph{NuSTAR} difference spectrum for intervals that lie above and below the best fit line, as in \citet{noda11}. If the scatter is caused by variations in the soft band, as expected if it is largely caused by absorption, then the difference spectrum over the \emph{NuSTAR} band would be dominated by the flux dependent variable component. Alternatively, if the variability was largely due to independent reflection variations then the difference spectrum should show strong reflection features. Finally, pivoting of the continuum would result in a difference spectrum well described by a less steep power law.

\begin{figure}[h!]
\centering
\includegraphics[width=\columnwidth]{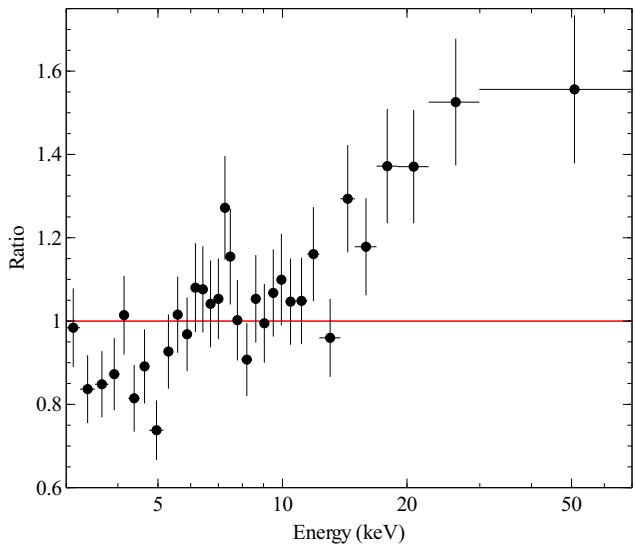}
\caption{\label{fluxratio}Ratio of the \emph{NuSTAR} difference spectrum between high and low flux spectra, calculated using flux-flux plots, to a $\Gamma=2$ power law, fit between 3--4 and 8--10 keV. FPMA and FPMB are grouped for plotting purposes, but fit separately.}
\end{figure}

Fig.~\ref{fluxratio} shows the ratio of the \emph{NuSTAR} difference spectrum from 3--70~keV to a $\Gamma=2$ power law. The data are binned to a minimum of 500 counts per bin, and the spectra from intervals below the best-fit line are used as the background for those above the line. We use the \emph{XMM-Newton} 0.5--2~keV band as a reference, and compare it with the \emph{NuSTAR} 10--50~keV band, to isolate any varying hard component. The figure shows a possible excess around $\sim7$~keV, and a prominent excess at high energies. 
The best-fitting powerlaw model gives a reduced chi-square of $\chi^2$/dof$=238/240=0.99$, but has a photon index of $1.75\pm0.03$, which is not consistent with the value obtained by fitting the full spectrum of MCG--06-30-15. Fixing $\Gamma$ at 2 gives a worse fit ($\chi^2$/dof$=299/241=1.24$) and cannot explain the excess high-energy flux, as shown in the figure. Fitting with pure reflection, using a simple \textsc{relconv*xillver} model with the parameters fixed at the best fit values for the reflection model (the reflection component in the absorption model is distant and should not vary fast enough to cause this scatter) and $\xi$ free to vary, gives an equivalent fit, but still not as good as the free powerlaw ($\chi^2$/dof$=296/241=1.23$). This means that the scatter is not predominantly due to independent reflection variability. 
Adding a $\Gamma=2$ power law to the reflection model results in a much improved fit, $\chi^2$/dof$=237/240=0.99$. This is equivalent to the best power law fit, leaving us with two possible scenarios that could cause the scatter in the flux-flux plots, absorption or pivoting of the primary power law (or both). In the first case, the variability all occurs at low energies due to effects such as the BLR occultations discussed in Section 6, so the high energy spectrum should have the same shape as the flux dependent variability (i.e. a power law plus blurred reflection, or partially covered power law plus distant reflection). In the second case, changes in the photon index of the primary power law should result in the high energy spectrum being well described by a more shallow power law.

We conclude that the scatter in the flux flux plots could either be due to pivoting of the primary continuum as found by Parker et al. 2014, or by absorption variability at low energies, or by a combination of the two.

\section{Conclusions}
We present results from a joint {\it NuSTAR} and {\it XMM-Newton} observational campaign of the bright Sy 1 galaxy MCG-6-30-15 and investigated the spectral variability of the source via a detailed time resolved analysis. The reflection scenario, where the primary variable power law continuum emission is reprocessed by the accretion disk, reproduces the data better than a scenario involving partial absorption by intervening structures. The former is preferred to the latter on statistical grounds, with a reduced $\chi^2$ of 1.10 versus 1.15, for about the same degrees of freedom ($\sim 4000$).

Our results can be summarized as follows:
\begin{itemize}
\item in the reflection scenario, the spectral variability can be either ascribed to a change of the ionization state of the disk or to an intrinsic change in the slope of the nuclear continuum, which is strongly favored on physical grounds (a variation of the photon index of the primary power law within $\Delta\Gamma\simeq0.3$). In the latter case the source is well described with gravitational light bending in the innermost regions of the accretion disk during the first part of the 2013 observational campaign and with intrinsic variations of the X-ray source in the latter part, this is in contrast to previous analyses \citep{fv03,lfm07,miniutti07};

\item the absorption model cannot account for all spectral variability if changes occur in the covering factor only. This is different than the behavior found in previous multi-epoch broad band analyses. A variation in the column density of the material along the line of sight is also needed, ranging between $10^{22}-10^{23}$ cm$^{-2}$;

\item we detected an occultation by a BLR cloud (N$_{\rm H}=2.2^{+0.8}_{-0.5}\times 10^{22}$ cm$^{-2}$) crossing the line of sight at a distance of $10^4$ R$_{\rm G}$, with a velocity of $v\simeq 3\times 10^3$ km s$^{-1}$ and a density of $n\simeq 7\times10^{9}$ cm$^{-3}$. This eclipsing event lasted for about 20 ks;

 \item using flux-flux plots we find strong correlations between {\it XMM} and {\it NuSTAR} energy bands, with an offset indicating a constant component at high energies. We identify significant variability uncorrelated with the source flux, manifested as significant scatter around the best-fit line in the flux-flux plots, too strong to be due to noise. We find that this variability could be caused either by pivoting of the primary power law or by changes in the absorption at low energies, or a combination of the two.
\end{itemize}

\section*{Acknowledgements}
AM acknowledges financial support from Fondazione Angelo Della Riccia.
AM and GM acknowledge financial support from Italian
Space Agency under grant ASI/INAF I/037/12/0-011/13. AM, GMatt, GMiniutti, ACF and EK acknowledge financial support
from the European Union Seventh Framework Programme
(FP7/2007-2013) under grant agreement n.312789. PA acknowledges financial support from Anillo ACT1101. This work was supported under NASA Contract No. NNG08FD60C, and made use of data from the {\it NuSTAR} mission, a project led by
the California Institute of Technology, managed by the Jet Propulsion
Laboratory, and funded by the National Aeronautics and Space
Administration. We thank the {\it NuSTAR} Operations, Software and
Calibration teams for support with the execution and analysis of
these observations.  This research has made use of the {\it NuSTAR}
Data Analysis Software (NuSTARDAS) jointly developed by the ASI
Science Data Center (ASDC, Italy) and the California Institute of
Technology (USA).

\bibliographystyle{apj}
\bibliography{sbs} 

\begin{thebibliography}{75}
\expandafter\ifx\csname natexlab\endcsname\relax\def\natexlab#1{#1}\fi

\bibitem[{{Arnaud}(1996)}]{xspec}
{Arnaud}, K.~A. 1996, in ASP Conf. Ser. 101: Astronomical Data Analysis
  Software and Systems V, 17

\bibitem[{{Ballantyne} {et~al.}(2003){Ballantyne}, {Vaughan}, \&
  {Fabian}}]{bvf03}
{Ballantyne}, D.~R., {Vaughan}, S., \& {Fabian}, A.~C. 2003, \mnras, 342, 239

\bibitem[{{Bennert} {et~al.}(2006){Bennert}, {Jungwiert}, {Komossa}, {Haas}, \&
  {Chini}}]{bennert06}
{Bennert}, N., {Jungwiert}, B., {Komossa}, S., {Haas}, M., \& {Chini}, R. 2006,
  \aap, 459, 55

\bibitem[{{Bianchi} {et~al.}(2009){Bianchi}, {Piconcelli}, {Chiaberge},
  {Bail{\'o}n}, {Matt}, \& {Fiore}}]{bianchi09c}
{Bianchi}, S., {Piconcelli}, E., {Chiaberge}, M., {Bail{\'o}n}, E.~J., {Matt},
  G., \& {Fiore}, F. 2009, \apj, 695, 781

\bibitem[{{Branduardi-Raymont} {et~al.}(2001){Branduardi-Raymont}, {Sako},
  {Kahn}, {Brinkman}, {Kaastra}, \& {Page}}]{bsk01}
{Branduardi-Raymont}, G., {Sako}, M., {Kahn}, S.~M., {Brinkman}, A.~C.,
  {Kaastra}, J.~S., \& {Page}, M.~J. 2001, \aap, 365, L140

\bibitem[{{Brenneman} \& {Reynolds}(2006)}]{br06}
{Brenneman}, L.~W. \& {Reynolds}, C.~S. 2006, \apj, 652, 1028

\bibitem[{{Cash}(1976)}]{cash76}
{Cash}, W. 1976, \aap, 52, 307

\bibitem[{{Chiang} \& {Fabian}(2011)}]{chfa11}
{Chiang}, C.-Y. \& {Fabian}, A.~C. 2011, \mnras, 414, 2345

\bibitem[{{Crummy} {et~al.}(2006){Crummy}, {Fabian}, {Gallo}, \&
  {Ross}}]{crummy06}
{Crummy}, J., {Fabian}, A.~C., {Gallo}, L., \& {Ross}, R.~R. 2006, \mnras, 365,
  1067

\bibitem[{{Dauser} {et~al.}(2013){Dauser}, {Garcia}, {Wilms}, {B{\"o}ck},
  {Brenneman}, {Falanga}, {Fukumura}, \& {Reynolds}}]{dauser13}
{Dauser}, T., {Garcia}, J., {Wilms}, J., {B{\"o}ck}, M., {Brenneman}, L.~W.,
  {Falanga}, M., {Fukumura}, K., \& {Reynolds}, C.~S. 2013, \mnras, 430, 1694

\bibitem[{{Dickey} \& {Lockman}(1990)}]{dl90}
{Dickey}, J.~M. \& {Lockman}, F.~J. 1990, \araa, 28, 215

\bibitem[{{Elvis} {et~al.}(2004){Elvis}, {Risaliti}, {Nicastro}, {Miller},
  {Fiore}, \& {Puccetti}}]{elvis04}
{Elvis}, M., {Risaliti}, G., {Nicastro}, F., {Miller}, J.~M., {Fiore}, F., \&
  {Puccetti}, S. 2004, \apjl, 615, L25

\bibitem[{{Fabian} \& {Ross}(2010)}]{faro10}
{Fabian}, A.~C. \& {Ross}, R.~R. 2010, \ssr, 157, 167

\bibitem[{{Fabian} \& {Vaughan}(2003)}]{fv03}
{Fabian}, A.~C. \& {Vaughan}, S. 2003, \mnras, 340, L28

\bibitem[{{Fabian} {et~al.}(2002){Fabian}, {Vaughan}, {Nandra}, {Iwasawa},
  {Ballantyne}, {Lee}, {De Rosa}, {Turner}, \& {Young}}]{fab02}
{Fabian}, A.~C., {Vaughan}, S., {Nandra}, K., {Iwasawa}, K., {Ballantyne},
  D.~R., {Lee}, J.~C., {De Rosa}, A., {Turner}, A., \& {Young}, A.~J. 2002,
  \mnras, 335, L1

\bibitem[{{Gabriel} {et~al.}(2004){Gabriel}, {Denby}, {Fyfe}, {Hoar}, {Ibarra},
  {Ojero}, {Osborne}, {Saxton}, {Lammers}, \& {Vacanti}}]{gabr04}
{Gabriel}, C., {Denby}, M., {Fyfe}, D.~J., {Hoar}, J., {Ibarra}, A., {Ojero},
  E., {Osborne}, J., {Saxton}, R.~D., {Lammers}, U., \& {Vacanti}, G. 2004, in
  Astronomical Society of the Pacific Conference Series, Vol. 314, Astronomical
  Data Analysis Software and Systems (ADASS) XIII, ed. {F.~Ochsenbein,
  M.~G.~Allen, \& D.~Egret}, 759--+

\bibitem[{{Garc{\'{\i}}a} {et~al.}(2013){Garc{\'{\i}}a}, {Dauser}, {Reynolds},
  {Kallman}, {McClintock}, {Wilms}, \& {Eikmann}}]{garcia13}
{Garc{\'{\i}}a}, J., {Dauser}, T., {Reynolds}, C.~S., {Kallman}, T.~R.,
  {McClintock}, J.~E., {Wilms}, J., \& {Eikmann}, W. 2013, \apj, 768, 146

\bibitem[{{George} \& {Fabian}(1991)}]{gf91}
{George}, I.~M. \& {Fabian}, A.~C. 1991, \mnras, 249, 352

\bibitem[{{Gierli{\'n}ski} \& {Done}(2004)}]{gido04}
{Gierli{\'n}ski}, M. \& {Done}, C. 2004, \mnras, 349, L7

\bibitem[{{Guainazzi} \& {Bianchi}(2007)}]{guabia07}
{Guainazzi}, M. \& {Bianchi}, S. 2007, \mnras, 374, 1290

\bibitem[{{Guainazzi} {et~al.}(1999){Guainazzi}, {Matt}, {Molendi}, {Orr},
  {Fiore}, {Grandi}, {Matteuzzi}, {Mineo}, {Perola}, {Parmar}, \&
  {Piro}}]{gmmo99}
{Guainazzi}, M., {Matt}, G., {Molendi}, S., {Orr}, A., {Fiore}, F., {Grandi},
  P., {Matteuzzi}, A., {Mineo}, T., {Perola}, G.~C., {Parmar}, A.~N., \&
  {Piro}, L. 1999, \aap, 341, L27

\bibitem[{{Harrison} {et~al.}(2013){Harrison}, {Craig}, {Christensen},
  {Hailey}, {Zhang}, {Boggs}, {Stern}, {Cook}, {Forster}, {Giommi},
  {Grefenstette}, {Kim}, {Kitaguchi}, {Koglin}, {Madsen}, {Mao}, {Miyasaka},
  {Mori}, {Perri}, {Pivovaroff}, {Puccetti}, {Rana}, {Westergaard}, {Willis},
  {Zoglauer}, {An}, {Bachetti}, {Barri{\`e}re}, {Bellm}, {Bhalerao},
  {Brejnholt}, {Fuerst}, {Liebe}, {Markwardt}, {Nynka}, {Vogel}, {Walton},
  {Wik}, {Alexander}, {Cominsky}, {Hornschemeier}, {Hornstrup}, {Kaspi},
  {Madejski}, {Matt}, {Molendi}, {Smith}, {Tomsick}, {Ajello}, {Ballantyne},
  {Balokovi{\'c}}, {Barret}, {Bauer}, {Blandford}, {Niel Brandt}, {Brenneman},
  {Chiang}, {Chakrabarty}, {Chenevez}, {Comastri}, {Dufour}, {Elvis}, {Fabian},
  {Farrah}, {Fryer}, {Gotthelf}, {Grindlay}, {Helfand}, {Krivonos}, {Meier},
  {Miller}, {Natalucci}, {Ogle}, {Ofek}, {Ptak}, {Reynolds}, {Rigby},
  {Tagliaferri}, {Thorsett}, {Treister}, \& {Urry}}]{nustar}
{Harrison}, F.~A., {Craig}, W.~W., {Christensen}, F.~E., {Hailey}, C.~J.,
  {Zhang}, W.~W., {Boggs}, S.~E., {Stern}, D., {Cook}, W.~R., {Forster}, K.,
  {Giommi}, P., {Grefenstette}, B.~W., {Kim}, Y., {Kitaguchi}, T., {Koglin},
  J.~E., {Madsen}, K.~K., {Mao}, P.~H., {Miyasaka}, H., {Mori}, K., {Perri},
  M., {Pivovaroff}, M.~J., {Puccetti}, S., {Rana}, V.~R., {Westergaard}, N.~J.,
  {Willis}, J., {Zoglauer}, A., {An}, H., {Bachetti}, M., {Barri{\`e}re},
  N.~M., {Bellm}, E.~C., {Bhalerao}, V., {Brejnholt}, N.~F., {Fuerst}, F.,
  {Liebe}, C.~C., {Markwardt}, C.~B., {Nynka}, M., {Vogel}, J.~K., {Walton},
  D.~J., {Wik}, D.~R., {Alexander}, D.~M., {Cominsky}, L.~R., {Hornschemeier},
  A.~E., {Hornstrup}, A., {Kaspi}, V.~M., {Madejski}, G.~M., {Matt}, G.,
  {Molendi}, S., {Smith}, D.~M., {Tomsick}, J.~A., {Ajello}, M., {Ballantyne},
  D.~R., {Balokovi{\'c}}, M., {Barret}, D., {Bauer}, F.~E., {Blandford}, R.~D.,
  {Niel Brandt}, W., {Brenneman}, L.~W., {Chiang}, J., {Chakrabarty}, D.,
  {Chenevez}, J., {Comastri}, A., {Dufour}, F., {Elvis}, M., {Fabian}, A.~C.,
  {Farrah}, D., {Fryer}, C.~L., {Gotthelf}, E.~V., {Grindlay}, J.~E.,
  {Helfand}, D.~J., {Krivonos}, R., {Meier}, D.~L., {Miller}, J.~M.,
  {Natalucci}, L., {Ogle}, P., {Ofek}, E.~O., {Ptak}, A., {Reynolds}, S.~P.,
  {Rigby}, J.~R., {Tagliaferri}, G., {Thorsett}, S.~E., {Treister}, E., \&
  {Urry}, C.~M. 2013, \apj, 770, 103

\bibitem[{{Iwasawa} {et~al.}(1996){Iwasawa}, {Fabian}, {Reynolds}, {Nandra},
  {Otani}, {Inoue}, {Hayashida}, {Brandt}, {Dotani}, {Kunieda}, {Matsuoka}, \&
  {Tanaka}}]{iwa96}
{Iwasawa}, K., {Fabian}, A.~C., {Reynolds}, C.~S., {Nandra}, K., {Otani}, C.,
  {Inoue}, H., {Hayashida}, K., {Brandt}, W.~N., {Dotani}, T., {Kunieda}, H.,
  {Matsuoka}, M., \& {Tanaka}, Y. 1996, \mnras, 282, 1038

\bibitem[{{Iwasawa} {et~al.}(1999){Iwasawa}, {Fabian}, {Young}, {Inoue}, \&
  {Matsumoto}}]{ify99}
{Iwasawa}, K., {Fabian}, A.~C., {Young}, A.~J., {Inoue}, H., \& {Matsumoto}, C.
  1999, \mnras, 306, L19

\bibitem[{{Jansen} {et~al.}(2001){Jansen}, {Lumb}, {Altieri}, {Clavel}, {Ehle},
  {Erd}, {Gabriel}, {Guainazzi}, {Gondoin}, {Much}, {Munoz}, {Santos},
  {Schartel}, {Texier}, \& {Vacanti}}]{xmm}
{Jansen}, F., {Lumb}, D., {Altieri}, B., {Clavel}, J., {Ehle}, M., {Erd}, C.,
  {Gabriel}, C., {Guainazzi}, M., {Gondoin}, P., {Much}, R., {Munoz}, R.,
  {Santos}, M., {Schartel}, N., {Texier}, D., \& {Vacanti}, G. 2001, \aap, 365,
  L1

\bibitem[{{Larsson} {et~al.}(2007){Larsson}, {Fabian}, {Miniutti}, \&
  {Ross}}]{lfm07}
{Larsson}, J., {Fabian}, A.~C., {Miniutti}, G., \& {Ross}, R.~R. 2007, \mnras,
  376, 348

\bibitem[{{Lee} {et~al.}(1999){Lee}, {Fabian}, {Brandt}, {Reynolds}, \&
  {Iwasawa}}]{lee99}
{Lee}, J.~C., {Fabian}, A.~C., {Brandt}, W.~N., {Reynolds}, C.~S., \&
  {Iwasawa}, K. 1999, \mnras, 310, 973

\bibitem[{{Lee} {et~al.}(2000){Lee}, {Fabian}, {Reynolds}, {Brandt}, \&
  {Iwasawa}}]{lfr00}
{Lee}, J.~C., {Fabian}, A.~C., {Reynolds}, C.~S., {Brandt}, W.~N., \&
  {Iwasawa}, K. 2000, \mnras, 318, 857

\bibitem[{{Lee} {et~al.}(2002){Lee}, {Iwasawa}, {Houck}, {Fabian}, {Marshall},
  \& {Canizares}}]{lih02}
{Lee}, J.~C., {Iwasawa}, K., {Houck}, J.~C., {Fabian}, A.~C., {Marshall},
  H.~L., \& {Canizares}, C.~R. 2002, \apjl, 570, L47

\bibitem[{{Lee} {et~al.}(2001){Lee}, {Ogle}, {Canizares}, {Marshall}, {Schulz},
  {Morales}, {Fabian}, \& {Iwasawa}}]{lee01}
{Lee}, J.~C., {Ogle}, P.~M., {Canizares}, C.~R., {Marshall}, H.~L., {Schulz},
  N.~S., {Morales}, R., {Fabian}, A.~C., \& {Iwasawa}, K. 2001, \apjl, 554, L13

\bibitem[{{Maiolino} {et~al.}(2010){Maiolino}, {Risaliti}, {Salvati},
  {Pietrini}, {Torricelli-Ciamponi}, {Elvis}, {Fabbiano}, {Braito}, \&
  {Reeves}}]{mrs10}
{Maiolino}, R., {Risaliti}, G., {Salvati}, M., {Pietrini}, P.,
  {Torricelli-Ciamponi}, G., {Elvis}, M., {Fabbiano}, G., {Braito}, V., \&
  {Reeves}, J. 2010, \aap, 517, A47

\bibitem[{{Martocchia} \& {Matt}(1996)}]{mama96}
{Martocchia}, A. \& {Matt}, G. 1996, \mnras, 282, L53

\bibitem[{{Matsumoto} {et~al.}(2003){Matsumoto}, {Inoue}, {Fabian}, \&
  {Iwasawa}}]{mif03}
{Matsumoto}, C., {Inoue}, H., {Fabian}, A.~C., \& {Iwasawa}, K. 2003, \pasj,
  55, 615

\bibitem[{{McHardy} {et~al.}(2005){McHardy}, {Gunn}, {Uttley}, \&
  {Goad}}]{mcha05}
{McHardy}, I.~M., {Gunn}, K.~F., {Uttley}, P., \& {Goad}, M.~R. 2005, \mnras,
  359, 1469

\bibitem[{{Miller} {et~al.}(2008){Miller}, {Turner}, \& {Reeves}}]{mtr08}
{Miller}, L., {Turner}, T.~J., \& {Reeves}, J.~N. 2008, \aap, 483, 437

\bibitem[{{Miller} {et~al.}(2009){Miller}, {Turner}, \& {Reeves}}]{mtr09}
---. 2009, \mnras, 399, L69

\bibitem[{{Miller} {et~al.}(2007){Miller}, {Turner}, {Reeves}, {George},
  {Kraemer}, \& {Wingert}}]{miller07}
{Miller}, L., {Turner}, T.~J., {Reeves}, J.~N., {George}, I.~M., {Kraemer},
  S.~B., \& {Wingert}, B. 2007, \aap, 463, 131

\bibitem[{{Miniutti} \& {Fabian}(2004)}]{mf04}
{Miniutti}, G. \& {Fabian}, A.~C. 2004, \mnras, 349, 1435

\bibitem[{{Miniutti} {et~al.}(2007){Miniutti}, {Fabian}, {Anabuki}, {Crummy},
  {Fukazawa}, {Gallo}, {Haba}, {Hayashida}, {Holt}, {Kunieda}, {Larsson},
  {Markowitz}, {Matsumoto}, {Ohno}, {Reeves}, {Takahashi}, {Tanaka},
  {Terashima}, {Torii}, {Ueda}, {Ushio}, {Watanabe}, {Yamauchi}, \&
  {Yaqoob}}]{miniutti07}
{Miniutti}, G., {Fabian}, A.~C., {Anabuki}, N., {Crummy}, J., {Fukazawa}, Y.,
  {Gallo}, L., {Haba}, Y., {Hayashida}, K., {Holt}, S., {Kunieda}, H.,
  {Larsson}, J., {Markowitz}, A., {Matsumoto}, C., {Ohno}, M., {Reeves}, J.~N.,
  {Takahashi}, T., {Tanaka}, Y., {Terashima}, Y., {Torii}, K., {Ueda}, Y.,
  {Ushio}, M., {Watanabe}, S., {Yamauchi}, M., \& {Yaqoob}, T. 2007, \pasj, 59,
  315

\bibitem[{{Miniutti} {et~al.}(2003){Miniutti}, {Fabian}, {Goyder}, \&
  {Lasenby}}]{min03}
{Miniutti}, G., {Fabian}, A.~C., {Goyder}, R., \& {Lasenby}, A.~N. 2003,
  \mnras, 344, L22

\bibitem[{{Miniutti} {et~al.}(2009){Miniutti}, {Ponti}, {Greene}, {Ho},
  {Fabian}, \& {Iwasawa}}]{mipogre09}
{Miniutti}, G., {Ponti}, G., {Greene}, J.~E., {Ho}, L.~C., {Fabian}, A.~C., \&
  {Iwasawa}, K. 2009, \mnras, 394, 443

\bibitem[{{Noda} {et~al.}(2011){Noda}, {Makishima}, {Uehara}, {Yamada}, \&
  {Nakazawa}}]{noda11}
{Noda}, H., {Makishima}, K., {Uehara}, Y., {Yamada}, S., \& {Nakazawa}, K.
  2011, \pasj, 63, 449

\bibitem[{{Otani} {et~al.}(1996){Otani}, {Kii}, {Reynolds}, {Fabian},
  {Iwasawa}, {Hayashida}, {Inoue}, {Kunieda}, {Makino}, {Matsuoka}, \&
  {Tanaka}}]{otani96}
{Otani}, C., {Kii}, T., {Reynolds}, C.~S., {Fabian}, A.~C., {Iwasawa}, K.,
  {Hayashida}, K., {Inoue}, H., {Kunieda}, H., {Makino}, F., {Matsuoka}, M., \&
  {Tanaka}, Y. 1996, \pasj, 48, 211

\bibitem[{{Parker} {et~al.}(2013){Parker}, {Marinucci}, {Brenneman}, {Fabian},
  {Kara}, {Matt}, \& {Walton}}]{parker13}
{Parker}, M.~L., {Marinucci}, A., {Brenneman}, L., {Fabian}, A.~C., {Kara}, E.,
  {Matt}, G., \& {Walton}, D.~J. 2013, ArXiv e-prints

\bibitem[{{Piconcelli} {et~al.}(2004){Piconcelli}, {Jimenez-Bail{\' o}n},
  {Guainazzi}, {Schartel}, {Rodr{\'{\i}}guez-Pascual}, \& {Santos-Lle{\'
  o}}}]{pico04}
{Piconcelli}, E., {Jimenez-Bail{\' o}n}, E., {Guainazzi}, M., {Schartel}, N.,
  {Rodr{\'{\i}}guez-Pascual}, P.~M., \& {Santos-Lle{\' o}}, M. 2004, \mnras,
  351, 161

\bibitem[{{Ponti} {et~al.}(2012){Ponti}, {Papadakis}, {Bianchi}, {Guainazzi},
  {Matt}, {Uttley}, \& {Bonilla}}]{ponti12}
{Ponti}, G., {Papadakis}, I., {Bianchi}, S., {Guainazzi}, M., {Matt}, G.,
  {Uttley}, P., \& {Bonilla}, N.~F. 2012, \aap, 542, A83

\bibitem[{{Puccetti} {et~al.}(2007){Puccetti}, {Fiore}, {Risaliti}, {Capalbi},
  {Elvis}, \& {Nicastro}}]{puc07}
{Puccetti}, S., {Fiore}, F., {Risaliti}, G., {Capalbi}, M., {Elvis}, M., \&
  {Nicastro}, F. 2007, \mnras, 377, 607

\bibitem[{{Reis} \& {Miller}(2013)}]{rm13}
{Reis}, R.~C. \& {Miller}, J.~M. 2013, \apjl, 769, L7

\bibitem[{{Reynolds} \& {Begelman}(1997)}]{rb97}
{Reynolds}, C.~S. \& {Begelman}, M.~C. 1997, \apj, 488, 109

\bibitem[{{Reynolds} {et~al.}(1997){Reynolds}, {Ward}, {Fabian}, \&
  {Celotti}}]{rey97b}
{Reynolds}, C.~S., {Ward}, M.~J., {Fabian}, A.~C., \& {Celotti}, A. 1997,
  \mnras, 291, 403

\bibitem[{{Risaliti} {et~al.}(2005){Risaliti}, {Elvis}, {Fabbiano}, {Baldi}, \&
  {Zezas}}]{ris05}
{Risaliti}, G., {Elvis}, M., {Fabbiano}, G., {Baldi}, A., \& {Zezas}, A. 2005,
  \apjl, 623, L93

\bibitem[{{Risaliti} {et~al.}(2007){Risaliti}, {Elvis}, {Fabbiano}, {Baldi},
  {Zezas}, \& {Salvati}}]{ris07}
{Risaliti}, G., {Elvis}, M., {Fabbiano}, G., {Baldi}, A., {Zezas}, A., \&
  {Salvati}, M. 2007, \apjl, 659, L111

\bibitem[{{Risaliti} {et~al.}(2002){Risaliti}, {Elvis}, \&
  {Nicastro}}]{risa02b}
{Risaliti}, G., {Elvis}, M., \& {Nicastro}, F. 2002, \apj, 571, 234

\bibitem[{{Risaliti} {et~al.}(2013){Risaliti}, {Harrison}, {Madsen}, {Walton},
  {Boggs}, {Christensen}, {Craig}, {Grefenstette}, {Hailey}, {Nardini},
  {Stern}, \& {Zhang}}]{risa13}
{Risaliti}, G., {Harrison}, F.~A., {Madsen}, K.~K., {Walton}, D.~J., {Boggs},
  S.~E., {Christensen}, F.~E., {Craig}, W.~W., {Grefenstette}, B.~W., {Hailey},
  C.~J., {Nardini}, E., {Stern}, D., \& {Zhang}, W.~W. 2013, \nat, 494, 449

\bibitem[{{Risaliti} {et~al.}(2009{\natexlab{a}}){Risaliti}, {Miniutti},
  {Elvis}, {Fabbiano}, {Salvati}, {Baldi}, {Braito}, {Bianchi}, {Matt},
  {Reeves}, {Soria}, \& {Zezas}}]{ris09}
{Risaliti}, G., {Miniutti}, G., {Elvis}, M., {Fabbiano}, G., {Salvati}, M.,
  {Baldi}, A., {Braito}, V., {Bianchi}, S., {Matt}, G., {Reeves}, J., {Soria},
  R., \& {Zezas}, A. 2009{\natexlab{a}}, \apj, 696, 160

\bibitem[{{Risaliti} {et~al.}(2011){Risaliti}, {Nardini}, {Elvis}, {Brenneman},
  \& {Salvati}}]{rne11}
{Risaliti}, G., {Nardini}, E., {Elvis}, M., {Brenneman}, L., \& {Salvati}, M.
  2011, \mnras, 417, 178

\bibitem[{{Risaliti} {et~al.}(2009{\natexlab{b}}){Risaliti}, {Young}, \&
  {Elvis}}]{rye09}
{Risaliti}, G., {Young}, M., \& {Elvis}, M. 2009{\natexlab{b}}, \apjl, 700, L6

\bibitem[{{Sako} {et~al.}(2003){Sako}, {Kahn}, {Branduardi-Raymont}, {Kaastra},
  {Brinkman}, {Page}, {Behar}, {Paerels}, {Kinkhabwala}, {Liedahl}, \& {den
  Herder}}]{sako03}
{Sako}, M., {Kahn}, S.~M., {Branduardi-Raymont}, G., {Kaastra}, J.~S.,
  {Brinkman}, A.~C., {Page}, M.~J., {Behar}, E., {Paerels}, F., {Kinkhabwala},
  A., {Liedahl}, D.~A., \& {den Herder}, J.~W. 2003, \apj, 596, 114

\bibitem[{{Sanfrutos} {et~al.}(2013){Sanfrutos}, {Miniutti},
  {Ag{\'{\i}}s-Gonz{\'a}lez}, {Fabian}, {Miller}, {Panessa}, \&
  {Zoghbi}}]{sanmi13}
{Sanfrutos}, M., {Miniutti}, G., {Ag{\'{\i}}s-Gonz{\'a}lez}, B., {Fabian},
  A.~C., {Miller}, J.~M., {Panessa}, F., \& {Zoghbi}, A. 2013, ArXiv e-prints

\bibitem[{{Shemmer} {et~al.}(2006){Shemmer}, {Brandt}, {Netzer}, {Maiolino}, \&
  {Kaspi}}]{shemmer06}
{Shemmer}, O., {Brandt}, W.~N., {Netzer}, H., {Maiolino}, R., \& {Kaspi}, S.
  2006, \apjl, 646, L29

\bibitem[{{Shih} {et~al.}(2002){Shih}, {Iwasawa}, \& {Fabian}}]{shih02}
{Shih}, D.~C., {Iwasawa}, K., \& {Fabian}, A.~C. 2002, \mnras, 333, 687

\bibitem[{{Smith} {et~al.}(2013){Smith}, {Guainazzi}, \& {Marinucci}}]{smith13}
{Smith}, M., {Guainazzi}, M., \& {Marinucci}, A. 2013, XMM-CCF-REL-300

\bibitem[{{Str{\"u}der} {et~al.}(2001){Str{\"u}der}, {Briel}, {Dennerl},
  {Hartmann}, {Kendziorra}, {Meidinger}, {Pfeffermann}, {Reppin}, {Aschenbach},
  {Bornemann}, {Br{\" a}uninger}, {Burkert}, \& {Elender}}]{struder01}
{Str{\"u}der}, L., {Briel}, U., {Dennerl}, K., {Hartmann}, R., {Kendziorra},
  E., {Meidinger}, N., {Pfeffermann}, E., {Reppin}, C., {Aschenbach}, B.,
  {Bornemann}, W., {Br{\" a}uninger}, H., {Burkert}, W., \& {Elender}, M. 2001,
  \aap, 365, L18

\bibitem[{{Tanaka} {et~al.}(1995){Tanaka}, {Nandra}, {Fabian}, {Inoue},
  {Otani}, {Dotani}, {Hayashida}, {Iwasawa}, {Kii}, {Kunieda}, {Makino}, \&
  {Matsuoka}}]{tanaka95}
{Tanaka}, Y., {Nandra}, K., {Fabian}, A.~C., {Inoue}, H., {Otani}, C.,
  {Dotani}, T., {Hayashida}, K., {Iwasawa}, K., {Kii}, T., {Kunieda}, H.,
  {Makino}, F., \& {Matsuoka}, M. 1995, \nat, 375, 659

\bibitem[{{Tatum} {et~al.}(2013){Tatum}, {Turner}, {Miller}, \&
  {Reeves}}]{tatum13}
{Tatum}, M.~M., {Turner}, T.~J., {Miller}, L., \& {Reeves}, J.~N. 2013, \apj,
  762, 80

\bibitem[{{Taylor} {et~al.}(2003){Taylor}, {Uttley}, \& {McHardy}}]{tum03}
{Taylor}, R.~D., {Uttley}, P., \& {McHardy}, I.~M. 2003, \mnras, 342, L31

\bibitem[{{Turner} {et~al.}(2004){Turner}, {Fabian}, {Lee}, \&
  {Vaughan}}]{tflv04}
{Turner}, A.~K., {Fabian}, A.~C., {Lee}, J.~C., \& {Vaughan}, S. 2004, \mnras,
  353, 319

\bibitem[{{Turner} {et~al.}(2003){Turner}, {Fabian}, {Vaughan}, \&
  {Lee}}]{tfv03}
{Turner}, A.~K., {Fabian}, A.~C., {Vaughan}, S., \& {Lee}, J.~C. 2003, \mnras,
  346, 833

\bibitem[{{Turner} {et~al.}(2001){Turner}, {Abbey}, {Arnaud}, {Balasini},
  {Barbera}, {Belsole}, {Bennie}, {Bernard}, {Bignami}, {Boer}, {Briel},
  {Butler}, {Cara}, {Chabaud}, {Cole}, {Collura}, {Conte}, {Cros}, \&
  {Denby}}]{turner01}
{Turner}, M.~J.~L., {Abbey}, A., {Arnaud}, M., {Balasini}, M., {Barbera}, M.,
  {Belsole}, E., {Bennie}, P.~J., {Bernard}, J.~P., {Bignami}, G.~F., {Boer},
  M., {Briel}, U., {Butler}, I., {Cara}, C., {Chabaud}, C., {Cole}, R.,
  {Collura}, A., {Conte}, M., {Cros}, A., \& {Denby}. 2001, \aap, 365, L27

\bibitem[{{Turner} {et~al.}(2007){Turner}, {Miller}, {Reeves}, \&
  {Kraemer}}]{turner07}
{Turner}, T.~J., {Miller}, L., {Reeves}, J.~N., \& {Kraemer}, S.~B. 2007, \aap,
  475, 121

\bibitem[{{Vaughan} \& {Edelson}(2001)}]{ve01}
{Vaughan}, S. \& {Edelson}, R. 2001, \apj, 548, 694

\bibitem[{{Vaughan} \& {Fabian}(2004)}]{vf04}
{Vaughan}, S. \& {Fabian}, A.~C. 2004, \mnras, 348, 1415

\bibitem[{{Walton} {et~al.}(2013){Walton}, {Nardini}, {Fabian}, {Gallo}, \&
  {Reis}}]{wana13}
{Walton}, D.~J., {Nardini}, E., {Fabian}, A.~C., {Gallo}, L.~C., \& {Reis},
  R.~C. 2013, \mnras, 428, 2901

\bibitem[{{Wilms} {et~al.}(2001){Wilms}, {Reynolds}, {Begelman}, {Reeves},
  {Molendi}, {Staubert}, \& {Kendziorra}}]{wilms01}
{Wilms}, J., {Reynolds}, C.~S., {Begelman}, M.~C., {Reeves}, J., {Molendi}, S.,
  {Staubert}, R., \& {Kendziorra}, E. 2001, \mnras, 328, L27

\bibitem[{{Yaqoob} \& {Padmanabhan}(2004)}]{yaq04}
{Yaqoob}, T. \& {Padmanabhan}, U. 2004, \apj, 604, 63

\bibitem[{{Young} {et~al.}(2005){Young}, {Lee}, {Fabian}, {Reynolds}, {Gibson},
  \& {Canizares}}]{ylf05}
{Young}, A.~J., {Lee}, J.~C., {Fabian}, A.~C., {Reynolds}, C.~S., {Gibson},
  R.~R., \& {Canizares}, C.~R. 2005, \apj, 631, 733

\end{thebibliography}
\newpage

\section*{APPENDIX A.1. Systematic uncertainties due to calibration effects in the EPIC-Pn energy scale reconstruction}

There is evidence that the calibration of the EPIC-pn energy scale in observations taken in 2013 is not as accurate as the nominal calibration goal ($\pm$10~eV), if they are reduced with the calibration files used in this paper. This effect is most likely due to inaccuracies in the long-term Charge Transfer Inefficiency (CTI) calibration (Smith et al. 2013\footnote[1]{available at {\tt http://xmm2.esac.esa.int/docs/documents/CAL-SRN-0300-1-0.pdf}}). More exactly, the CTI should be over-corrected in recent observations. In this Appendix we discuss how such a calibration inaccuracy affects the astrophysical results discussed in this paper.

In order to quantify the inaccuracy of the energy scale, we used the "line-like" feature at $\simeq$2.3~keV present in the residuals against all the fits discussed in this paper. We interpret this feature as due to inaccuracies of the energy scale, which are the most apparent at the energy where the gradient of the effective area is the steepest. An alternative interpretation of the same feature in terms of local inaccuracies of the effective area calibration requires implausibly large deviations of the optics Gold coating reflection law from the physically-motivated models used in the effective area calibration. We modified the Pulse Invariant (PI) column of the calibrated event list generated by {\it epproc} in steps of one PI unit in the range [-100:100]; extracted a time-average spectrum from each of these modified event lists using the same procedure as described in Sect.~2 of this paper; and calculated the $\chi^2$ when fitting each of these spectra with a simple power-law in the 1.5--5~keV energy band. The shift minimizing the $\chi^2$ is $\Delta PI = +8 \pm 2$ (1$\sigma$error; we remind that one PI unit corresponds to approximately 5~eV). This result is consistent with the energy of the Mn K$_{\alpha,1}$ line (laboratory energy: 5.8876~keV) measured in a long calibration observation taken close to the astrophysical observations discussed in this paper (Obs.\#0411781301, October 25 2012): the difference against the laboratory energy was $\Delta E_{MnK} = 31 \pm 4$~eV. The inaccuracy of the energy scale is consistent with being energy-independent within the statistical errors.

\begin{table}[h!]
\centering
\caption{Systematic error on the best-fit parameters in Tab.~3 induced by the CTI overcorrection in EPIC-pn spectra}
\label{tab1mg}
\begin{tabular}{lc} \hline \hline
$\Delta \log(\xi_1)$ & $\pm 0.02$ \\
$\Delta N_{H_1}$ (10$^{22}$~cm$^{-2}$) & $^{+0.15}_{-0.015}$ \\
$\Delta \log(\xi_2)$ & $^{+0.05}_{-0.22}$ \\
$\Delta N_{H_2}$ (10$^{22}$~cm$^{-2}$) & $^{+0.10}_{-0.011}$ \\
$\Delta \log(\xi_{refl})$ & $<0.01$ \\
$\Delta \log(N_{Fe})$ & $0.2$ \\
$\Delta \log(Z_{Fe})$ & $<0.1$ \\
$\Delta q$ & 0.18 \\
$\Delta i$ & 8 \\
$\Delta a$ & $<0.01$ \\
$\Delta \Gamma$ & 0.014 \\ \hline \hline
\end{tabular}
\end{table}

We then modified the combined EPIC-pn event list through a Monte-Carlo algorithm, changing the PI column according to a Gaussian distribution with average +40~eV and standard deviation 10~eV. From this modified event list we extracted spectra in HR-resolved intervals as discussed in Sect.~4, and repeated the fits using the complete best-fit model therein discussed. The difference between the best-fit parameters measured on spectra extracted from the standard, and from the Monte-Carlo modified event lists are shown in Tab.~\ref{tab1mg}. They are lower, or at most comparable to the statistical errors in Tab.~3. This demonstrates that the inaccuracy of the energy scale affecting the EPIC-pn data discussed in this paper does not dominate the error budget of our analysis. However, systematic uncertainties cannot be neglected altogether. In particular, the warm absorber solutions required for Intervals\#5 and \#11 are significantly different from those in Tab.~3 (cf. Tab.~\ref{tab2mg})\\

\begin{table*}[h!]
\centering
\caption{Difference in warm absorber best-fit parameters for Intervals\#5 and \#11.}
\label{tab2mg}
\begin{tabular}{lcccc} \hline \hline
Interval & $\Delta \log(\xi_1)$ & $\Delta N_{H_1}$ (10$^{22}$~cm$^{-2}$) & $\Delta \log(\xi_2)$ & $\Delta N_{H_2}$ (10$^{22}$~cm$^{-2}$) \\ \hline
5 & 0.04 & -0.2 & 0.7 & 0.13 \\
11 & 0.04 & 0.7 & -1.6 & -0.2 \\ \hline \hline
\end{tabular}
\end{table*}

\end{document}